\documentclass[12pt,a4paper]{article}
\pdfoutput=1
\usepackage{jheppub}
\usepackage{amsmath}
\usepackage{amsfonts}
\usepackage{amssymb}
\usepackage{graphicx}
\usepackage{mathrsfs}
\usepackage{comment}
\usepackage{bbold}
\usepackage{slashed}
\usepackage{xcolor}

\newcommand{\bone}{1\!\! 1}
\newcommand{\e}{\textrm{e}}

\newcommand{\tr}{\textrm{tr}}

\newcommand{\Det}{\textrm{det}}
\newcommand{\Detp}[1]{\Det{}^{#1}}
\newcommand{\Zz}{\mathcal{Z}}
\newcommand{\Zzp}{\mathcal{Z}^{\prime}}
\newcommand{\ps}{\slashed{p}}
\newcommand{\symb}{\textrm{symb}^{-1}}
\newcommand{\non}{\nonumber\\}
\newcommand{\cZ}{\Zz}

\newcommand{\ppm}{{p^{-}}}

\newcommand{\gpr}{g_{\parallel}}
\newcommand{\gpp}{g_{\perp}}
\newcommand{\Fh}{\hat{F}}
\newcommand{\eBm}{\left(\frac{e B}{m^{2}}\right)}

\def\Eins{{\mathchoice {\rm 1\mskip-4mu l} {\rm 1\mskip-4mu l}
{\rm 1\mskip-4.5mu l} {\rm 1\mskip-5mu l}}}
\newcommand{\mn}{\mu\nu}
\def\bear{\begin{eqnarray}}
\def\ear{\end{eqnarray}}
\def\e{{\rm e}}
\def\half{\frac{1}{2}}
\def\coth{{\rm coth}}%% extra brakets needed to stop latex switching font
\def\tanh{{\rm tanh}}%% extra brakets needed to stop latex switching font
\def\sinh{{\rm sinh}} %% extra brakets needed to stop latex switching font

\newcommand{\ud}{\mathrm{d}}
\newcommand{\LCm}{{\scriptscriptstyle -}} %LC supersripts
\newcommand{\LCp}{{\scriptscriptstyle +}}
\newcommand{\LCpm}{{\scriptscriptstyle \pm}}

\newcommand{\LCperp}{{\scriptscriptstyle \perp}}

\newcommand{\Anton}[1]{{\color{blue}#1}}

\newcommand{\be}{\begin{equation}}
\newcommand{\ee}{\end{equation}}
\newcommand{\bi}{\begin{itemize}}
\newcommand{\ei}{\end{itemize}}
\newcommand{\bea}{\begin{eqnarray}}
\newcommand{\eea}{\end{eqnarray}}

\def\tr{{\rm tr}}
\centerline{\title{Reducible contributions to quantum electrodynamics in external fields.}}

\author[a,b]{Naser  Ahmadiniaz,}
\author[c]{James P. Edwards,}
\author[d]{Anton Ilderton}

\affiliation[a]{Helmholtz-Zentrum Dresden-Rossendorf, Bautzner Landstra\ss e 400, 01328 Dresden, Germany}
\affiliation[b]{Center for Relativistic Laser Science, Institute for Basic Science, 61005 Gwangju, Korea}
\affiliation[c]{Instituto de F\'isica y Matem\'aticas
Universidad Michoacana de San Nicol\'as de Hidalgo
Edificio C-3, Apdo. Postal 2-82
C.P. 58040, Morelia, Michoac\'an, M\'exico}
\affiliation[d]{Centre for Mathematical Sciences, University of Plymouth, PL48AA, UK}
\emailAdd{n.ahmadiniaz@hzdr.de}
\emailAdd{jedwards@ifm.umich.mx}
\emailAdd{anton.ilderton@plymouth.ac.uk}
\abstract{We consider one-particle reducible (1PR) contributions to QED and scalar QED processes in external fields, at one-loop and two-loop order. We investigate three cases in detail: constant crossed fields, constant magnetic fields, and plane waves. We find that 1PR tadpole contributions in plane waves and constant crossed fields are non-zero, but contribute only divergences to be renormalised away. In constant magnetic fields, on the other hand, tadpole contributions give physical corrections to processes at one loop and beyond. Our calculations are exact in the external fields and we give strong and weak field expansions in the magnetic case.}

\begin{document}
\maketitle
% !TEX root = manuscript.tex
\section{Introduction}
Quantum field theory in the presence of an external field is a rich area of physics that finds applications in heavy ion collisions, accelerator physics, astrophysical scenarios and intense laser-particle physics. If the field is strong, then it must be treated without recourse to perturbation theory in the coupling to the background field, making such instances of great theoretical and phenomenological interest. This is possible if the field configuration is simple, or highly symmetric.
%yielding fertile ground for theoretical analysis.

This area of field theory was pioneered by Euler and Heisenberg who, taking a \textit{constant} electromagnetic background, calculated the one-loop effective Lagrangian for QED~\cite{EHL} (see the calculations of Schwinger and Weisskopf \cite{WEH, Schwinger} for the corresponding calculations in scalar QED, and~\cite{Dunne:2004nc} for a review of these results).  As is well known, one physical consequence revealed by the Euler-Heisenberg Lagrangian (EHL) is the instability of the vacuum to the application of strong electric fields, which leads to particle / anti-particle pair creation (the Schwinger mechanism). This effect has recently received renewed attention~\cite{Dunne:2008kc,Bell,Fedotov,Bulanov,Gonoskov} due to the prospects of investigating pair creation using future laser facilities. For the status of current and future laser facilities, making study of these backgrounds of great experimental interest for the coming years see, for example, the information at \cite{cilex,corels,eli,xfel}).

Related results now exist for the effective action at two- and three loops \cite{dunne2000two,huet2018three} in a constant background, (anti-)self-dual backgrounds \cite{SD1, SD2, SD3} and at one-loop order for various non-constant backgrounds such as Sauter pulses \cite{Schneider:2014mla,Torgrimsson:2017pzs,Torgrimsson:2017cyb} 
and a pulsed Hermite and Laguerre-Gaussian laser beam \cite{PhysRevD.96.116004}. See also \cite{PhysRevD.87.125020} for the full mass range analysis of the QED effective action for a nontrivial background with some special symmetry. These have been used to study low energy photon amplitudes \cite{Low1, Low2} and the structure of the quantum vacuum, see~\cite{King:2015tba} for a recent review. Aside from this, the particle propagator can also be constructed exactly (non-perturbatively) in the presence of constant fields, plane waves, and other symmetric fields, allowing the calculation of a variety of electron-seeded and photon-seeded processes, see~\cite{RitusRev,DiPiazza:2011tq,King:2015tba,Seipt:2017ckc} for reviews.

\begin{figure}
\centering
	\includegraphics[width=0.3\textwidth]{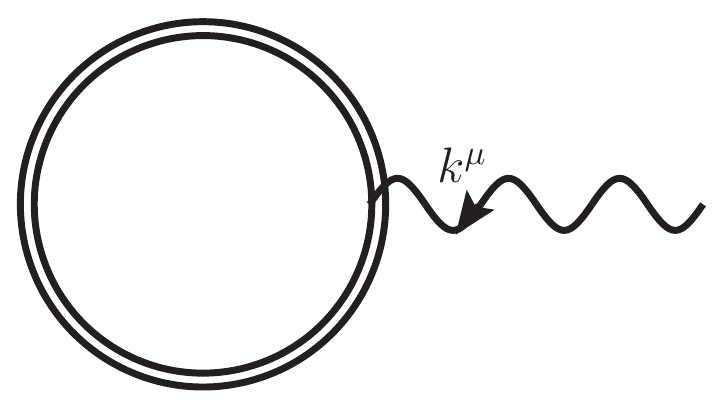}
	\caption{\label{figTad} The ``tadpole'' diagram formally vanishes by momentum conservation (in vacuum it vanishes by Furry's theorem) but can contribute when sewn to a larger diagram. The double line indicates the particle propagator dressed to all orders by the background field.}
\end{figure}
Recently, however, it was found that historical calculations had overlooked the possibility of one particle reducible (1PR) contributions to processes in constant background fields~\cite{Gies:2016yaa,Karbstein:2017pbf,karbstein2017tadpole}. These contributions involve a tadpole, displayed in figure \ref{figTad},  attached somewhere in the corresponding Feynman diagram describing the process. The tadpole is linear in the exchanged (off-shell) photon momentum, $k^{\mu}$, and momentum conservation implies that it can be supported only for $k^{\mu} = 0$. This may seem to suggest that the tadpole contribution vanishes, which has long been asserted in this area of quantum field theory \cite{HolgerBook, DitReu}. However, the propagator joining the tadpole to the remainder of the diagram diverges at $k^\mu=0$, and a careful analysis shows that a finite result remains. For example, joining two tadpoles in any covariant gauge (in the following we use Feynman gauge) leads to a momentum integral of the form
\be\label{SEW-1}
	\int\! \ud^{D}k \, \delta^{D}(k) \frac{k^{\mu}k^{\nu}}{k^{2}} = \frac{1}{D}\eta^{\mu\nu},
\ee
where the tensor structure of the right hand side is determined entirely by covariance. This result is the origin of surviving contributions from reducible diagrams and we shall appeal to it in our analysis to come below. 

\begin{figure}[t!]
\centering
	\includegraphics[width=0.4\textwidth]{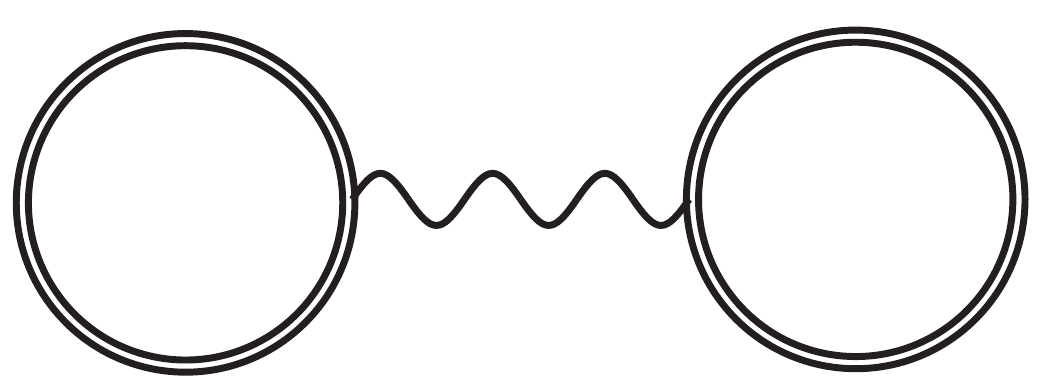}
	\caption{\label{figDumb} The 1PR contribution to the two-loop EHL: the ``dumbell'' diagram, consisting of two tadpoles sewn together. The double lines indicate the particle propagators dressed to all orders by the background field.}
\end{figure}
\begin{figure}
\centering
	\includegraphics[width=0.4\textwidth]{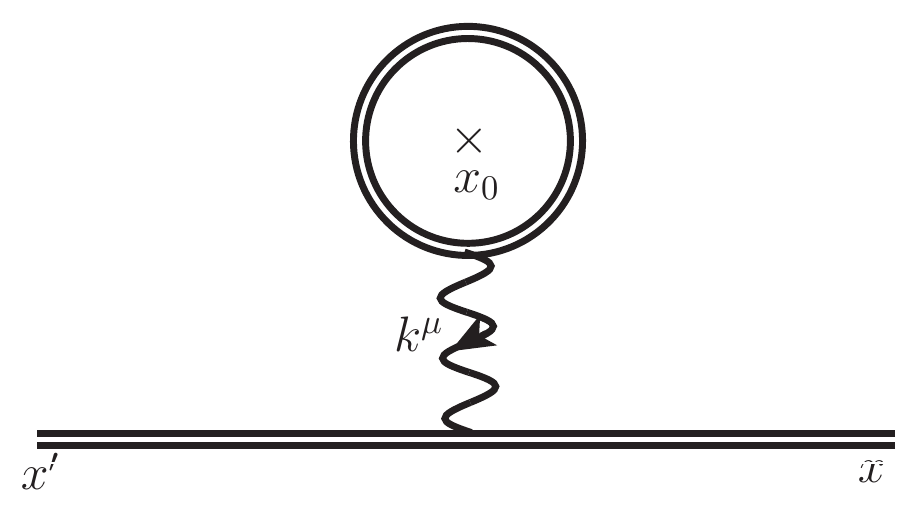}
	\caption{The 1PR contribution to the particle self-energy, where a tadpoles is sewn to the propagator (shown in position space; $x_{0}$ is the centre of mass of the loop.).  The double lines indicate the particle propagators dressed to all orders by the background field.}
\label{figProp}
\end{figure}

The original discovery \cite{Gies:2016yaa} focussed on the reducible contribution to the \textit{two}-loop QED Euler-Heisenberg Lagrangian (the ``dumbbell'' of figure \ref{figDumb}), which should be added to the original irreducible diagram consisting of a virtual photon exchanged in a single loop. This was rapidly extended to scalar QED \cite{GKUs1}, where it was then found that that there were additional reducible corrections to the scalar propagator in a constant background even at \textit{one-loop} order. The results were further developed to an analogous result for the spinor propagator in \cite{GKUs2} (see~\cite{HolgerBook,mckeon1994radiative,ScalProp,ahmadiniaz2016multiphoton} for the tree level propagators). These processes are shown in figure \ref{figProp}; they are of the same order in coupling as the usual irreducible one-loop contributions to the particle self-energies.

In the cases of both the two-loop EHL and one-loop self-energy corrections there are covariant formulae expressing the reducible contributions in terms of derivatives of lower order objects. For the two-loop EHL, the reducible contribution can be written as
\begin{equation}
	\mathcal{L}^{(2)1PR}[F] = \frac{\partial \mathcal{L}^{(1)}}{\partial F^{\mu\nu}} \frac{\partial \mathcal{L}^{(1)}}{\partial F_{\mu\nu}}\,,
	\label{GKEH}
\end{equation}
where $\mathcal{L}^{(1)}[F]$ is the one-loop EHL and $F$ is the field strength tensor of the background field. This is valid for spinor and scalar QED upon use of the appropriate EHL. For the one-loop propagator, choosing Fock-Schwinger gauge \cite{FockFS, SchwingerFS} for the background field centered at one of the endpoints of the line, the momentum space version of the covariant formula for spinor matter is
\begin{equation}
	S^{(1) 1PR} (p | F)= \frac{\partial S}{\partial F^{\mu\nu}}\frac{\partial \mathcal{L}^{(1)}}{\partial F_{\mu\nu}} \;,
	\label{GKProp}
\end{equation}
where now $ S(p | F)$ is the tree level propagator in the constant background. % (for the scalar case we replace all spinor propagators, $S$, by their scalar counterparts, $D$).

Although these equations are complete for arbitrary constant fields, further insight can be gained by examining these contributions for some specific field configurations, where both general features and field-specific phenomena can be seen. The EHL is of phenomenological interest for its relation to pair creation, vacuum birefringence (photon helicity flip) etc., in the strong fields of intense laser experiments~\cite{Heinzl:2006xc} or astrophysical scenarios, and of theoretical importance due to the AAM conjecture~\cite{AAM}. Similarly, loop corrections to the electron propagator, including, in general, 1PI contributions, contribute to the Ritus mass shift~\cite{RitusShift}, and to $g-2$ in the presence of a background field \cite{MuExt}. Clearly, then, it is important to know what the physical consequences of the new reducible contributions are, since their effects have been missed since the earliest days of background field QED. In this paper we therefore determine the reducible contributions for some background fields of phenomenological interest and analyse the results.

This paper is organised as follows. In section \ref{secCrossed} we show that the 1PR contribution to the EHL in crossed fields vanishes, whilst the tadpole correction to the electron self-energy picks up an additional renormalisation from the background field. In section \ref{secBField} we turn to a constant magnetic field. Here both the EHL and the electron propagator pick up a \textit{finite}, physical part from the reducible diagrams that cannot be absorbed into renormalisation. Finally in section \ref{secPlane} we consider a non-constant background, namely an arbitrary plane wave. These fields are of central importance to the modelling of intense laser experiments in the relativistic and quantum regimes (and also provide a smooth limit to the constant crossed field case). If 1PR contributions had been overlooked it would have significant implications for several existing literature calculations. This would include electron spin flip at one loop~\cite{Meuren:2011hv} and loop corrections to photon emission which are essential for the correct modelling of radiation reaction~\cite{Ilderton:2013dba}, a topic of current experimental investigation~\cite{Cole:2017zca,Poder:2018ifi}. Due to the many symmetries of plane waves, we are, notably, able to calculate the tadpole correction to \textit{any} diagram. We show that this only contributes terms which can be removed by renormalisation. We conclude and discuss our results in section \ref{secConc}.

Throughout, we present details for spinor QED in the main text, and then state the corresponding scalar QED (sQED) results, relegating the details of the scalar calculations to the appendix.

% !TEX root = manuscript.tex
\section{Crossed electric and magnetic fields}
\label{secCrossed}
Whilst the covariant formulae (\ref{GKProp})--(\ref{GKEH}) are  compact, we require more explicit expressions for the derivatives involved in order to calculate the form of the reducible contributions for a chosen background, so we begin there. In this section we give the general 1PR contribution to the QED EHL (at two loops) and to the electron propagator (at one loop) in an arbitrary constant field, and then specialise to the case of constant crossed fields, where a simple argument shows that the 1PR diagrams correspond to renormalisation.

\subsection{Explicit 1PR contributions}

A convenient representation of the 1-loop EHL is the ``proper-time'' representation derived in the worldline (or first quantised) approach to QED \cite{Strass1, ChrisRev, BG0, UsRep} and dating back to Schwinger \cite{schwinger1951gauge}. The (un-renormalised) EHL for spinor matter coupled to a constant electromagnetic background admits the proper-time integral representation (in Minkowski spacetime) \cite{Sp1, Dittrich:2000wz, BG0, Shaisultanov:1995tm}
\begin{equation}
	\mathcal{L}^{(1)}[F] = -2\int_{0}^{\infty}\frac{\ud s}{s}(4\pi i s)^{-\frac{D}{2}} \e^{-im^{2}s} \Detp{-\frac{1}{2}}\Big[\frac{ \tanh\Zz}{\Zz} \Big]\,,
	\label{L1WL}
\end{equation}
where $\mathcal{Z}_{\mu\nu} := e s F_{\mu\nu}$ with $F_{\mu\nu}$ the constant field strength tensor for the background. Likewise, the spinor propagator in a constant background field has the compact integral representation presented in \cite{GKUs2} (based upon the results of \cite{SpinProp})
\begin{equation}
	S(p | F) = \int_{0}^{\infty}\ud s\, \big[i(m - \ps) + i \gamma \cdot \tanh \Zz \cdot p\big]\e^{-is\left( m^{2} + p \cdot \frac{\tanh \Zz}{\Zz} \cdot p\right)}\symb \Big\{\e^{-\frac{1}{4}\eta \cdot \tanh \Zz \cdot \eta }\Big\}\,,
	\label{SpWL}
\end{equation}
where the ``symbol map''  is defined by 
\begin{equation}
\textrm{symb} 
\bigl(\hat\gamma^{[\alpha\beta\cdots\rho]}\bigr) \equiv 
\eta^\alpha\eta^\beta\ldots\eta^\rho\,,
\label{defsymb}
\end{equation}
with $\hat\gamma^{\mu} \equiv i\sqrt{2} \gamma^{\mu}$ and where $ \hat\gamma^{[\alpha\beta\cdots\rho]}$ denotes the totally anti-symmeterised product,
\begin{align}
\hat\gamma^{[\alpha_1\alpha_2\cdots \alpha_n]} \equiv \frac{1}{n!}\sum_{\pi\in S_n} {\rm sign}(\pi) \hat\gamma^{\alpha_{\pi(1)}}\hat\gamma^{\alpha_{\pi(2)}} \cdots \hat\gamma^{\alpha_{\pi(n)}} \, .
\label{Defantisymm}
\end{align}
Explicitly 
\begin{equation}
	{\rm symb}^{-1}\Big[\e^{-\frac{1}{4}\eta\cdot \tanh\cZ\cdot\eta}\Big] = \bone+\frac{1}{4}\cZ^{\mu\nu}[\gamma^\mu,\gamma^\nu]-\frac{1}{8}\epsilon^{\mu\nu\alpha\beta}\cZ_{\mu\nu}\cZ_{\alpha\beta}\gamma_5. 
	\label{InvSymb}
\end{equation}
These results follow from recent advances in treating tree level processes within the worldline formalism.

Applying the formulae (\ref{GKEH}) and (\ref{GKProp}) we arrive at the results of \cite{GKUs1, GKUs2} for the 1PR contribution to the EHL
\begin{align}
\hspace{-2.5em}	\mathcal{L}^{(2)1PR} = -&\frac{4 e^{2}}{D}\int_{0}^{\infty} \ud s (4\pi i s)^{-\frac{D}{2}}\e^{-i m^{2} s} \int_{0}^{\infty}\ud s' (4\pi i s')^{-\frac{D}{2}}\e^{-im^{2} s^{\prime}} \nonumber \\
\hspace{-2.5em}	&\times\Detp{-\frac{1}{2}}\Big[\frac{\tanh \Zz}{\Zz} \Big] \Detp{-\frac{1}{2}}\Big[\frac{\tanh \Zz'}{\Zz'} \Big] \tr \left[\left(\mathcal{\dot{G}}_{B} - \mathcal{G}_{F}\right) \cdot\left(\mathcal{\dot{G}}'_{B} - \mathcal{G}'_{F}\right)\right]\,,
	\label{L2}
\end{align}
where $\mathcal{G}_{B}$ and $\mathcal{G}_{F}$ are the coincidence limits of the bosonic and fermionic ``worldline Green functions'' in the presence of the constant background field\footnote{All functions of the matrix $\Zz$ are defined by their power series, all of which involve only non-negative powers of the same matrix.},
\begin{equation}
\mathcal{\dot{G}}_{B} = \coth \Zz - \frac{1}{\Zz}, \quad \qquad \mathcal{G}_{F} = \tanh \Zz\, ,
\end{equation} 
and to the self-energy,
\begin{align}\label{S1}
	&S^{(1)1PR}(p)=-e^{2}\int_{0}^{\infty} \ud s \,s\,\e^{-is(m^{2} + p \cdot \frac{\tanh \Zz}{\Zz} \cdot p)}\int_{0}^{\infty}\ud s^{\prime}(4 \pi i s^{\prime})^{-\frac{D}{2}}\e^{-im^{2}s^{\prime}}
\textrm{det}^{-\frac{1}{2}}\bigg[\frac{\tanh \Zzp}{\Zzp} \bigg]\nonumber \\
& \bigg\lbrace\bigg[ i(m - \ps) + i\gamma \cdot \tanh \Zz \cdot p \bigg]
\bigg[	- s\, p \cdot \frac{\Zz - \sinh \Zz \cdot \cosh \Zz}{\Zz^{2} \cdot \cosh^{2}\Zz} \cdot \Xi^{\prime} \cdot p -i  \Xi^{\prime}_{\mn}\frac{\partial}{\partial {\cal Z}_{\mn}} \bigg] 
\nonumber\\
&\hspace{2cm}+ \gamma \cdot {\rm sech}^{2}\Zz \cdot
\Xi^{\prime} \cdot p 
\bigg\rbrace
\textrm{symb}^{-1} \bigg\{ \e^{-\frac{1}{4}\eta \cdot \tanh \Zz \cdot \eta}\bigg\}\,,
\end{align}
where 
\begin{equation}
\Xi[F] \equiv  -i\Big[\frac{1}{\Zz} -  \frac{1}{\sinh \Zz \cdot \cosh \Zz} \Big].
\label{defXi}
\end{equation}
All primed variables in the above equations refer to the proper time parameter, $ \Zzp := eFs^{\prime}$. We now evaluate these contributions in the special case of crossed electric and magnetic fields of equal strength.

\subsection{Constant crossed fields}
We consider the class of constant fields with vanishing Maxwell invariants, $F_{\mu\nu}F^{\mu\nu} = 4(|\mathbf{E}|^{2} - |\mathbf{B}|^{2}) = 0$ and $F_{\mu\nu}\widetilde{F}^{\mu\nu} = 4\mathbf{E}\cdot \mathbf{B} = 0$, where the dual field strength tensor is defined as usual by $\widetilde{F}^{\mu\nu} := \frac{1}{2}\epsilon^{\mu\nu\alpha\beta}F_{\alpha\beta}$. Furthermore $F^{3}_{\mu\nu} \equiv F_{\mu\alpha} F^{\alpha\beta} F_{\beta\nu} = 0$ for such fields and all higher powers also vanish.

Note then, that as there are no invariants which can be built from the field alone, the EHL for crossed fields must be independent of the background field, i.e.~is effectively zero. To see that the 1PR contribution at two-loop order respects this, we note that for all constant backgrounds the one-loop EHL, $\mathcal{L}^{(1)}$, is an even function of the field strength tensor meaning that its derivative with respect to $F$ is odd. Given that for the crossed field background $F^{3}$ vanishes, it is clear that the factor $\frac{\partial \mathcal{L}^{(1)}}{\partial F^{\mu\nu}}$ is linear in $F$ (recall that although the one-loop EHL reduces to a ($D = 4$ divergent) field -independent constant for crossed fields, one should take the derivative of $\mathcal{L}^{(1)}$ for an arbitrary background before specialising the result to the crossed field case).  Consequently the crossed field tadpole, when attached to any diagram will be linear in the coupling of the tadpole's loop to the background field.  For this reason we can immediately deduce that the 1PR contributions to the one-loop self-energy and the two-loop EHL can be absorbed by renormalisation (we discuss this below). This general argument applies to both spinor and scalar QED\footnote{Thanks go to Christian Schubert for helpful discussions on these points.}.

It is useful for the studies below of \textit{nontrivial} cases to see how the above result appear through the covariant formulae (\ref{GKEH}) and (\ref{GKProp}). This also allows us to determine the exact coefficient of the part linear in the background. For constant crossed fields we may always choose coordinates such that the field strength tensor
\begin{equation}
	F_{\mu\nu} = \begin{pmatrix}0&B&0&0\\-B&0&0&B\\0&0&0&0\\0&-B&0&0\end{pmatrix}.
\end{equation}
As can be checked, $F^3_{\mu\nu} = 0$, so that all hyperbolic trigonometric functions that enter the proper time representations of the general 1PR contributions, above, are at most quadratic in $\Zz$ and $\Zzp$.

\begin{figure}
\centering
	\includegraphics[width=0.9\textwidth]{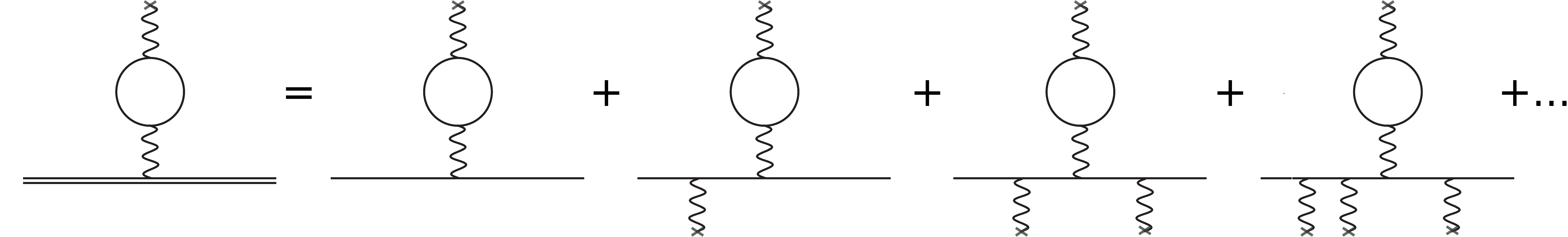}
	\caption{The schematic diagrammatic expansion of the 1PR contribution to the self energy interactions with the background field for crossed fields. Being linear in the coupling of the loop, there is only one low energy photon attached thereto.}
\label{figPropExpSp}
\end{figure}

For the 1PR contribution to the self-energy, evaluating the trigonometric functions in (\ref{S1}) and computing the $s'$ integral leads to the representation (the super-script minus refers to light-cone coordinates, $x^{\pm} := x^{0} \pm x^{3}$ and square brackets indicate anti-symmetrisation of indices without a combinatorical factor)
\begin{align}
\label{Scd}	&S^{(1)1PR} = \frac{e^{2}}{m^{2}}\left(\frac{m^{2}}{4 \pi}\right)^{\frac{D}{2}} \frac{eB}{m^{2}}\Gamma\Big[2 - \frac{D}{2}\Big]\int_{0}^{\infty} \ud s \,s\,\e^{-is\left(p^{2} + m^{2} + \frac{z^{2}}{3} \ppm^{2}\right)} \\
& \left[ \frac{1}{2}\Big\{ i\left(m - \ps\right) \, , \, \frac{4is}{9} z \ppm^{2} - \frac{2}{3} \gamma^{-}\gamma^{1} \Big\} + iz\gamma^{[-}p^{1]} \left(\frac{4is}{9} z \ppm^{2} - \frac{2}{3} \gamma^{-}\gamma^{1}\right)\right] \big[\Eins + z\gamma^{-}\gamma^{1} \big] \nonumber\;,
\end{align}
in which $\{\cdot, \cdot \}$ denotes the anticommutator, and $z = eBs$. The leading $\frac{eB}{m^2}$ arises from the integral over the loop proper time $s^{\prime}$ and, as argued above, the result is linear in this coupling of the loop to the crossed field background (this is because $\Xi'$ is linear in $\Zzp$ and it enters every term of the integrand). As such we see that this 1PR contribution can be absorbed simply by an additional (infinite, in $D = 4$) renormalisation of the photon propagator, as shown in figure \ref{figPropExpSp}. It therefore has no physical significance, once the photon propagator has been correctly renormalised. The result (\ref{Scd}) is suitable for numerical integration and is amenable to an expansion in the background field, but as it corresponds to renormalisation, it is not necessary to pursue that here.

The Feynman diagrams corresponding to an expansion of (\ref{Scd}) in powers of the coupling to the background field are shown in figure \ref{figPropExpSp}. External photon legs with a cross correspond to the background field and have vanishingly small energy. Since the loop couples linearly to the background, only one such low energy photon is attached to it, whereas the line couples to an arbitrary number of photons. We discuss the specific form of the vertices at the end of the next subsection. 

For the 1PR correction to the EHL, (\ref{L2}), it is sufficient to note that for crossed fields 
\begin{equation}
\label{GbGfXd}
	\mathcal{\dot{G}}_{B} = \frac{1}{3}\Zz \;, \quad \mathcal{G}_{F} = \Zz	\;, \quad \implies \mathcal{\dot{G}}_{B} - \mathcal{G}_{F} = -\frac{2\Zz}{3}\,,
\end{equation}
so that the relative contribution of spin is $-3$ times that in scalar QED. Hence the integrand in (\ref{L2}) contains the factor, linear in $\Zzp$ as expected,
\begin{equation}
	 \tr \left[\left(\mathcal{\dot{G}}_{B} - \mathcal{G}_{F}\right) \cdot\left(\mathcal{\dot{G}}'_{B} - \mathcal{G}'_{F}\right)\right]  = \frac{4}{9} \tr \Big[\Zz \cdot \Zzp\Big].
\end{equation}
However $\tr(F^{2}) = 0$, being the first of the Maxwell invariants. The remaining parts of the integrand of (\ref{L2}) are field independent, so that the integrand identically vanishes. Thus we recover the result that there is no 1PR correction (not even additional renormalisation) to the two-loop spinor EHL for constant crossed fields.

\subsection{Scalar QED}
For scalar QED the story is much the same. The 1PR contribution requires the proper-time representation of the scalar one-loop EHL given in the appendix and evaluates to\footnote{The spinor and scalar results are related by the so-called ``replacement rules'' discussed in \cite{ChrisRev} which roughly amounts to replacing products of $\mathcal{\dot{G}}_{B}$ by the same product minus its counterpart with $\mathcal{\dot{G}}_{B} \rightarrow \mathcal{G}_{F}$ and an overall change of normalisation of the path integral. We again work with the un-renormalised EHL.} 
\begin{align}\label{L2sc}
	\mathcal{L}^{(2)1PR} = -\frac{ e^{2}}{D}&\int_{0}^{\infty} \!\ud s (4\pi i s)^{-\frac{D}{2}}\e^{-im^{2} s}\int_{0}^{\infty}\!\ud s' (4\pi i s')^{-\frac{D}{2}}\e^{-im^{2} s^{\prime}} \nonumber \\
	&\times\Detp{-\frac{1}{2}}\Big[\frac{\sinh \Zz}{\Zz} \Big] \Detp{-\frac{1}{2}}\Big[\frac{\sinh \Zz'}{\Zz'} \Big]  \tr \big[\mathcal{\dot{G}}_{B} \cdot \mathcal{\dot{G}}'_{B}\big]\,.
\end{align}
The expansion of $\mathcal{\dot{G}}_{B}$ for crossed fields is proportional to $\Zz$ as for the spinor case as it is an odd function, $\mathcal{\dot{G}}_{B} = \Zz/3$, so that as above the integrand is proportional to $\tr [\Zz \cdot \Zzp] = 0$. Once again, the 1PR contribution to the 2-loop scalar EHL is zero for constant crossed fields.

For the one-loop correction to the propagator the proper time representation of the tree level propagator in the appendix leads to the explicit form 
\begin{align}
	D^{(1)1PR}(p) = \frac{e^2}{2} &\int_0^\infty \ud s'(4\pi is')^{-\frac{D}{2}}\e^{-im^2s'}{\rm det}^{-\half}\Big[\frac{\sinh \cZ^{\prime}}{\cZ^{\prime}}\Big] \nonumber \\
	&\times\int_0^\infty \ud s\,s\,\e^{-im^2 s}{\rm det}^{-\half}\Big[\cosh\cZ\Big]\e^{-is\,p\cdot\frac{\tanh \cZ}{\cZ}\cdot p}\non
&\times \Big[s\,p\cdot\frac{\sinh \cZ\cdot\cosh\cZ-\cZ}{\cZ^2\cdot\cosh^2\cZ}\cdot\dot{\mathcal{G'}}_{B}\cdot p+\half\tr\Big(i\tanh\cZ\cdot\dot{\mathcal{G'}}_B\Big)\Big]\,.\non
\label{D1sc}
\end{align}
Evaluating this for crossed fields gives
\begin{equation}
\hspace{-1em}	D^{(1)1PR}(p)=\frac{e^2\ppm^2}{9m^{2}}\left(\frac{m^{2}}{4\pi}\right)^{\frac{D}{2}}\eBm \Gamma\Big[ 2 - \frac{D}{2}\Big]\int_0^\infty \ud s\,s^2\, \e^{-is[p^2+ m^{2}+  \frac{z^2}{3}\ppm^2]}\,z.
\label{D1X}
\end{equation}
This is, of course, once again linear in the coupling of the loop to the background field (the leading factor of $\eBm$), and as such it again corresponds to a trivial renormalisation. The remaining integral with respect to $s$ is finite in $D = 4$.

\begin{figure}
\centering
	\includegraphics[width=0.6\textwidth]{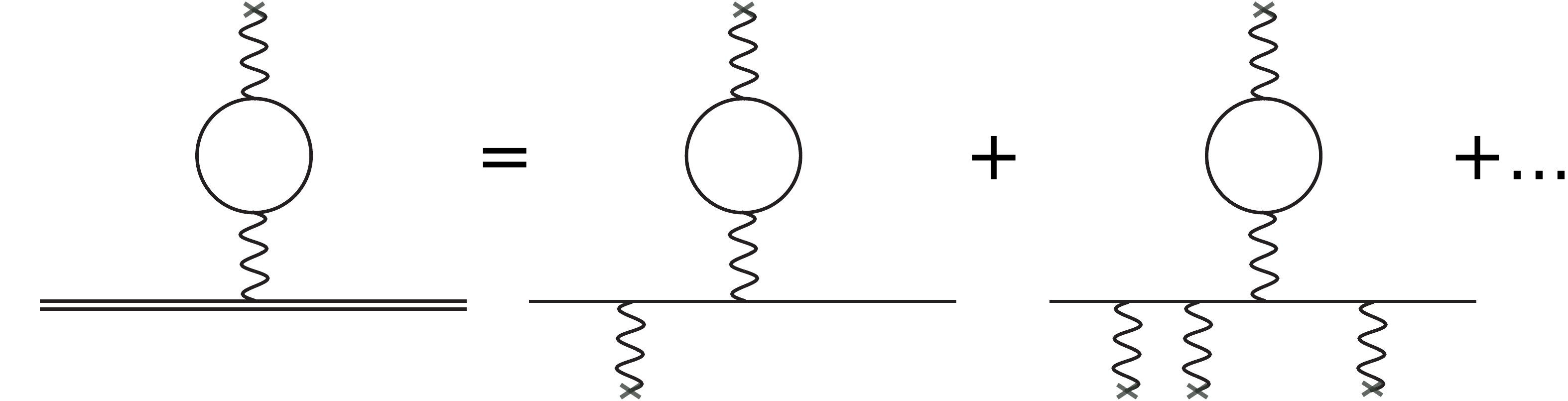}
	\caption{The expansion of the 1PR contribution to the scalar propagator in the crossed field case. Note that in Fock-Schwinger gauge there is an odd number of free low energy photons of the background coupled to the line, and one low energy photon attached to the loop as before.}
\label{figPropExpSc}
\end{figure}

Let us briefly compare the spinor and scalar cases for the self-energy, (\ref{Scd}) and (\ref{D1X}) to examine some general features. Although both are linear in the loop's coupling to the background field, the spinor case begins at zeroth order in the coupling of the background to the line ($z^{0}$), whereas the scalar result, being an odd function of this coupling, begins at order $z$. This would seem to miss a contribution from the vacuum propagator and one low energy photon attached to the loop (first diagram in figure \ref{figPropExpSp}) but this is an artefact of Fock-Schwinger gauge. Since the tree level scalar propagator in the constant background, (\ref{DpWl}), is an even function of $\Zz$ in this gauge an expansion in powers of $\Zz$ will produce insertions of an even number of low energy photons. Then (\ref{GKProp}) implies connecting one of these photons with a photon from the expansion of the loop, leaving an odd number of free photons remaining on the line (there can therefore be no contribution to (\ref{D1X}) at $\mathcal{O}(z^{0})$, for example). This expansion is shown in figure \ref{figPropExpSc} for the scalar propagator.

In the second order formulation \cite{SO1, SO2} of spinor QED, however, which the worldline formalism is based upon, there is an additional vertex beyond the 3-point and seagull vertices of scalar QED. This extra 3-point vertex couples the spin degrees of freedom to the background in a gauge invariant way\footnote{Its Feynman rule is an insertion $e\sigma^{\mu\nu}k_{\nu}$ where $\sigma^{\mu\nu}$ are the spin 1/2 generators of the Lorentz group and $k$ is the external photon momentum.} and contributes to processes with an arbitrary number of photon insertions. As such, (\ref{SpWL}) is neither even nor odd in $\Zz$ and so its expansion involves arbitrary powers of this variable. Thus (\ref{Scd}) involves terms constant and linear in the coupling of the line to the background field and hence its expansion in figure \ref{figPropExpSp}.

After this warm up where we have verified the general argument that the crossed field tadpole affects only the renormalisation required during quantisation, we turn to a more interesting field configuration where the 1PR contributions imply physical corrections.

% !TEX root = manuscript.tex
\section{Constant magnetic field}
\label{secBField}
In this section we consider constant fields with Maxwell invariants $\mathcal{F}<0$, $\mathcal{G}=0$, in contrast to the above. In the current case it is always possible to choose a frame such that the background is a pure magnetic field pointing along the $z$-direction, say, so that $\vec{B}=B\hat{z}$. Calculating the 1PR contributions to the EHL and propagators for scalar and spinor QED in a magnetic field, we will see that there is a physical contribution, beyond renormalisation. We will also explicitly compute the result in the weak field approximation.

In this background the only non-vanishing components of the field strength tensor are $F_{12}=-B$ and $F_{21}=B$
\begin{equation}
F_{\mu\nu} =
\left(
\begin{array}{*{4}{c}}
0&0&0&0\\
0&0&-B&0\\
0&B&0&0\\
0&0&0&0
\end{array}
\right)\,.
\label{b-field}
\end{equation} 
In this section it will be convenient to make use of the following projection matrices:
\bear
\hat{F} = 
\begin{pmatrix}
  0 & 0 & 0 & 0 \\
  0&0 & -1 & 0 \\
  0&1&0& 0  \\
  0& 0 & 0 & 0 
 \end{pmatrix} \;,
 \qquad
 g_{\perp}= 
\begin{pmatrix}
  0& 0 & 0 & 0 \\
  0&1& 0 & 0 \\
  0&0&1& 0  \\
  0& 0 & 0 & 0 
 \end{pmatrix} \;,
\qquad
  g_{\parallel}= 
\begin{pmatrix}
  -1 & 0 & 0 & 0 \\
  0&0 & 0 & 0 \\
  0&0&0& 0  \\
  0& 0 & 0 & 1 
 \end{pmatrix} \;,
\ear 
which will help us to write the expansions of the trigonometric functions and determinants in (\ref{L2}) and (\ref{S1}). For example, the determinant factor can be simplified as 
\begin{align}
\Detp{-\frac{1}{2}}\Bigl\lbrack\frac{\tanh \Zz}{\Zz}\Big]&= \Detp{-\frac{1}{2}}\Bigl\lbrack \gpr + \frac{\tan z}{z} \gpp \Big] = \frac{z}{\tan z}\,,
\end{align}
where $z=eBs$. 

%%%%%%%%%
\subsection{1PR contribution to the two loop EHL}
%%%%%%%%%
%
For the 1PR contribution to the two-loop EHL we also need the result
\begin{equation}
	\dot{\mathcal{G}}_{B} - \mathcal{G}_{F} =-\left( \cot z - \frac{1}{z}  + \tan z \right)\Fh \;.
\end{equation}
Defining, $\mathcal{J}(z) = (z/\tan z) (\cot z -1/z+\tan z )$, (\ref{L2}) can be written in this background as
\be
\label{L2B}
\mathcal{L}^{(2)1PR} = \frac{4 e^{2}}{D}\int_{0}^{\infty} \!\ud s (4\pi i s)^{-\frac{D}{2} } \e^{-im^2 s}\mathcal{J}(z) \int_{0}^{\infty}\!\ud s' (4\pi i s')^{-\frac{D}{2}}\e^{-im^2 s'} \mathcal{J}(z') \;.
\ee
%\be
%\begin{split}
%\label{L2B}
%\mathcal{L}^{(2)1PR} = \frac{4 e^{2}}{D}\int_{0}^{\infty}& dT (4\pi i T)^{-\frac{D}{2} }\frac{z}{\tan z} \left( \cot z - \frac{1}{z}  + \tan z \right) \\
%&\int_{0}^{\infty}dT' (4\pi i T')^{-\frac{D}{2}}\frac{z'}{\tan z'} \left( \cot z' - \frac{1}{z'}  + \tan z' \right)
%\end{split}
%\ee
%
As the integrand contains arbitrary positive powers of $z$ and $z'$ it is clear that this contribution to the EHL cannot be absorbed by renormalisation, so that this represents an important physical correction at two-loop order.

\begin{figure}
\centering
	\includegraphics[width=0.95\textwidth]{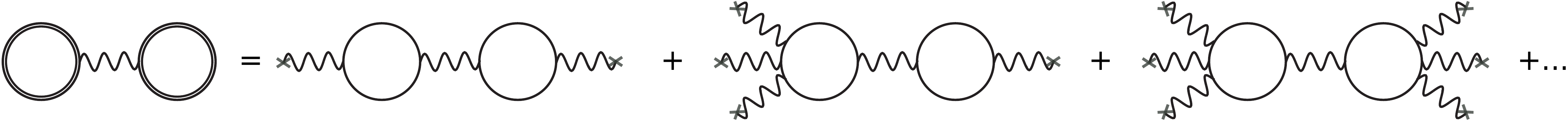}
	\caption{The weak field expansion of the reducible contribution to the two-loop spinor EHL. The single photons attached to loops are subtracted by renormalisation. Note that Furry's theorem (now that the propagators are single-lined, vacuum propagators) implies that only an odd number of low energy photons can be attached to each loop.}
\label{figL2B}
\end{figure}

For weak fields  (that is, $B/B_{\rm cr}\ll1$ with $B_{cr}=m^2/e\simeq 4.41\times 10^{13} {\rm G}$ the critical field strength) we can expand the integrand in order to determine explicitly the first non-trivial contribution. Using $\mathcal{J}(z) = \frac{2z}{3} + \frac{4z^{3}}{45} + \mathcal{O}(z^{5})$ we find
%$\frac{z}{\tan z} \left( \cot z - \frac{1}{z}  + \tan z \right) = \frac{2z}{3} + \frac{4z^{3}}{45} + \mathcal{O}(z^{5})$ we get
\begin{equation}\label{L2B2}
		\mathcal{L}^{(2)1PR} = \frac{4 e^{2}}{D m^{4}}\left(\frac{m^{2}}{4\pi}\right)^{D}\left[\frac{2}{3}\eBm \Gamma\Big[2 - \frac{D}{2}\Big] - \frac{4}{45}\eBm^{3} \Gamma\Big[4 - \frac{D}{2}\Big] + \ldots \right]^{2}.
\end{equation}
The first term in the large square bracket is divergent in $D = 4$, but is linear in the coupling of the respective loop to the background, so can be removed by renormalisation as we saw in the case of crossed fields (in $D = 4$ it is sufficient to replace $J(z) \rightarrow J(z) -\frac{2}{3}z$ in a similar spirit to the renormalisation of the one-loop EHL). The higher order terms, though, are physical and start at order $\eBm^{6}$. This should be contrasted with the weak field expansion of the irreducible contribution to the two-loop EHL \cite{RitusEHL} that starts at order $\eBm^{4}$. We show the expansion of (\ref{L2B}) in figure \ref{figL2B}; the two factors in large square brackets of (\ref{L2B2}) correspond to the first and second diagrams in the expansion shown in the figure.

It is also very interesting to consider the strong field limit, since it has recently been shown that the strong field asymptotic limit of the EHL is determined, at all loop orders, by one-loop \textit{reducible} contributions~\cite{Karbstein:2019wmj}. Here we will confirm the leading-order behaviour argued by \cite{Karbstein:2019wmj} at two loops by explicitly evaluating the integrals in (\ref {L2B}). Since there are two copies of the same integral it suffices to focus on one, and then square the result. Returning to Euclidean space-time by replacing $s\rightarrow -iT$, therefore $z\rightarrow -iz$, we must also work with the renormalised Lagrangian \cite{GD1}. Anticipating our eventual specialisation to $D = 4$, the renormalisation proceeds as mentioned above, where the pole in $s$ is subtracted so that the integral we require is
\bear
\mathcal{I}&=&-\int\limits_0^\infty\!\ud T\, (4\pi T)^{-\frac{D}{2}}\,\e^{-m^2T}\Big[{\rm coth} z-z\, {\rm csch}^2 - \frac{2}{3}z\Big]  \\
&=&-\frac{1}{eB}\Big(\frac{4\pi}{eB}\Big)^{-\frac{D}{2}}\int\limits_0^\infty\!\ud z \,\e^{-\frac{m^2}{eB}z}  z^{-\frac{D}{2}} \Big[{\rm coth} z-z\,{\rm csch}^2 z- \frac{2}{3}z\Big]=-\frac{1}{eB}\Big(\frac{4\pi}{eB}\Big)^{-\frac{D}{2}}(\mathcal{I}_1-\mathcal{I}_2 - \mathcal{I}_{3})\,.\nonumber
\label{i1i2}
  \ear
The complete integral is now finite in $D = 4$. We begin with $\mathcal{I}_1$; this is just the Laplace transform $F(\omega)$ of the function $f(z)$ where
\bear
	f(z)=z^{-\frac{D}{2}}\,{\rm coth}z \; \quad \text{and} \quad \omega=\frac{m^2}{eB}\,.
\ear
The Laplace transform can be expressed in terms of the Hurwitz zeta function $\zeta[x,q]$ as
\bear
\mathcal{I}_1=2^{\frac{D}{2}-1}\Gamma\Big[1-\frac{D}{2}\big] \Big(\zeta\big[1-\frac{D}{2},\frac{m^2}{2eB}\Big]+\zeta\big[1-\frac{D}{2},1+\frac{m^2}{2eB}\big]\Big)\,.
\ear
Likewise we observe that the integral $\mathcal{I}_{2}$ may be expressed by introducing an auxiliary parameter, $\alpha$, as
\bear
	\mathcal{I}_2 = \int\limits_0^\infty\!\ud z\,\e^{-\frac{m^2}{eB}z}\,z^{1-\frac{D}{2}}\,{\rm csch}^2z
		= -\frac{\partial}{\partial\alpha}\int\limits_0^\infty\!\ud z\,\e^{-\frac{m^2}{eB}z}\,z^{-\frac{D}{2}}\,{\rm coth}(\alpha z)\Big\vert_{\alpha=1}\,.
\ear
Making the change of variables $z \rightarrow \frac{z}{\alpha}$ we learn that $\mathcal{I}_{2} = -\frac{\partial}{\partial \alpha} \alpha^{\frac{D}{2} - 1}F(\frac{\omega}{\alpha}) \big|_{\alpha = 1}$ so we may re-use the result for $\mathcal{I}_{1}$ to get
\bear
\mathcal{I}_2&=& 2^{\frac{D}{2}-1}\Gamma\Big[2-\frac{D}{2}\Big]\bigg\{ \zeta\big[1-\frac{D}{2},\frac{m^2}{2eB}\big]+\zeta\big[1-\frac{D}{2},1+\frac{m^2}{2eB}\big] \\
&&\hspace{3.3cm}-\frac{m^2}{2eB}\Big(\zeta\big[2-\frac{D}{2},\frac{m^2}{2eB}\big]+\zeta\big[2-\frac{D}{2},1+\frac{m^2}{2eB}\big]\Big)\bigg\}\, . \nonumber
\ear 
Finally the integral $\mathcal{I}_{3}$ is trivial,
\begin{equation}
	\mathcal{I}_{3} = \frac{2}{3}\eBm^{2-\frac{D}{2}}\Gamma\Big[2 - \frac{D}{2}\Big].
\end{equation}
Substituting these two expressions into (\ref{i1i2}) and setting $D = 4-2\epsilon$, the $\frac{1}{\epsilon}$ divergence in $\mathcal{I}_{3}$ cancels that in $\mathcal{I}_{1} - \mathcal{I}_{2}$. The remaining, finite, expression can then be expanded for large $B$ to obtain the strong field expansion. One easily finds that $\mathcal{I} \sim \frac{eB}{24 \pi^{2}} \ln \eBm$ and so  the leading order strong field behaviour of the reducible two-loop contribution to the EHL is
\bear
\mathcal{L}^{(2)1PR} \sim \frac{1}{2} B^2 \left[ \alpha \beta_{1} \ln\eBm \right]^2\,
\ear
where $\beta_{1} = \frac{1}{3\pi}$ is the order $\alpha$ coefficient of the $\beta$-function in spinor QED. This correctly reproduces the results presented in~\cite{Karbstein:2019wmj} at two-loop order. 

Note also that the asymptotic behaviour can be read off from the finite (in $\epsilon$) contribution of the renormalisation term in (\ref{i1i2}):
\begin{align}
	\frac{1}{eB} \left( \frac{4\pi}{eB} \right)^{-2} \mathcal{I}_{3} (D = 4 - 2\epsilon)\bigg|_{\mathcal{O}(\epsilon^{0})} &= \frac{2}{3 eB}\left(\frac{4\pi}{eB}\right)^{-2}\eBm^{\epsilon}\Gamma[\epsilon]\bigg|_{\mathcal{O}(\epsilon^{0})} \nonumber \\
	&\sim \frac{eB}{24 \pi^{2}}\ln \eBm
\end{align}
where the behaviour holds asymptotically (the expansion of the prefactor to the integral in $\epsilon$ contributes subleading field-dependent terms that are killed by $\mathcal{I}_{1}$ and $\mathcal{I}_{2}$ as with the $\frac{1}{\epsilon}$ poles discussed above). This connection between the strong field asymptotic behaviour and the renormalisation term introduced to render the proper time integral finite (that also give the $\beta$-function coefficient) is well known at one-loop order \cite{GD1, GD2}.

%%%%%%%%%
\subsection{1PR contribution to the self energy}
%%%%%%%%%
As for the 1PR contribution to the spinor self-energy, after plugging the field strength tensor in (\ref{b-field}) into (\ref{S1}) and using the matrices defined above one gets the following expressions for the required terms: 
%%%%
\bear
p \cdot \frac{\tanh\Zz }{\Zz} \cdot p &=& p_{\parallel}^{2} + \frac{\tan z}{z} p_{\perp}^{2}\,,  \\
i(m- \ps) + i \gamma \cdot \tanh \cZ \cdot p&=&i(m-\slashed{p})+i{\rm tan}z\, \gamma^{[2}p^{1]} \,,\\
\Xi' &=& i\Big[\frac{1}{z'} -  \frac{1}{\sin z'  \cos z'} \Big]\hat{F}\,,\\
-ip \cdot \frac{\Zz - \sinh \Zz \cdot \cosh \Zz}{\Zz^{2} \cdot \cosh^{2}\Zz} \cdot \Xi^{\prime} \cdot p 
&=&\Big[\frac{\sec^2z}{z}\!-\!\frac{\tan z}{z^2}\Big]\Big[\frac{1}{z'}\!-\!\frac{1}{\sin z'\cos z'}\Big]p_{\perp}^{2}\,,~~ \\
\gamma \cdot {\rm sech}^{2}\Zz \cdot\Xi^{\prime} \cdot p
&=& i\sec^2 z\Big[\frac{1}{z'} -  \frac{1}{\sin z'  \cos z'} \Big]\gamma^{[2}p^{1]}\,,\\
\textrm{symb}^{-1} \bigg\{\e^{-\frac{1}{4}\eta \cdot \tanh \Zz \cdot \eta}\bigg\} &=& \Eins + \frac{1}{2} \tan z \, \gamma^{[2} \gamma^{1]}\,, \\
-i\Xi'_{\mu\nu}\frac{\partial}{\partial \cZ_{\mu\nu}}\textrm{symb}^{-1} \bigg\{\e^{-\frac{1}{4}\eta \cdot \tanh \Zz \cdot \eta}\bigg\}
&=& \frac{1}{2}\left[\frac{1}{z'} - \frac{1}{\sin z' \cos z'}\right] \sec^{2}z \, \gamma^{[2} \gamma^{1]} \,.
\ear
We note that the final term (and its derivative) from the inverse symbol map, (\ref{InvSymb}) vanishes for a constant magnetic field and have used $\gamma \cdot \hat{F} \cdot \gamma = [\gamma^{2}, \gamma^{1}]$ and $\gamma\cdot \hat{F} \cdot p = \gamma^{[2}p^{1]}$. Using these results we find that the one-loop 1PR correction to the electron propagator in a constant magnetic field is given by 
\begin{align}\label{igen}
	&S^{(1)1PR}(p) = -ie^{2}\int_{0}^{\infty} \!\ud s s\,\e^{-is(m^{2}+p_\parallel^2 + \frac{\tan z}{z}p_\perp^{2})}\int_{0}^{\infty}\!\ud s^{\prime}(4i \pi s^{\prime})^{-\frac{D}{2}}\e^{-im^{2}s^{\prime}}
\big(\cot z'-z'\csc^2 z'\big) \nonumber \\
&\hspace{0.7cm}\times\bigg\lbrace\Big[-is p_{\perp}^{2}\Big(\frac{\sec^2z}{z}-\frac{\tan z}{z^2}\Big)\big( m-\slashed{p}+{\rm tan}z\, \gamma^{[2}p^{1]}\big) +\sec^2 z\, \gamma^{[2}p^{1]} \Big] \Big[\Eins+\frac{1}{2}\tan z [\gamma^{2}, \gamma^{1}]\Big] \nonumber \\
 &\hspace{4cm}+\frac{1}{2}(m-\slashed{p}+\tan z\,\gamma^{[2}p^{1]})\sec^{2}z[\gamma^{2}, \gamma^{1}]\bigg\} \;.
\end{align}
This general result is non-vanishing, as we show directly below and generalises easily to a constant magnetic field in an arbitrary direction. The integrand involves arbitrary powers of $z'$, so cannot be completely absorbed by renormalisation.

\begin{figure} [t]
	\centering
	\includegraphics[width=0.9\textwidth]{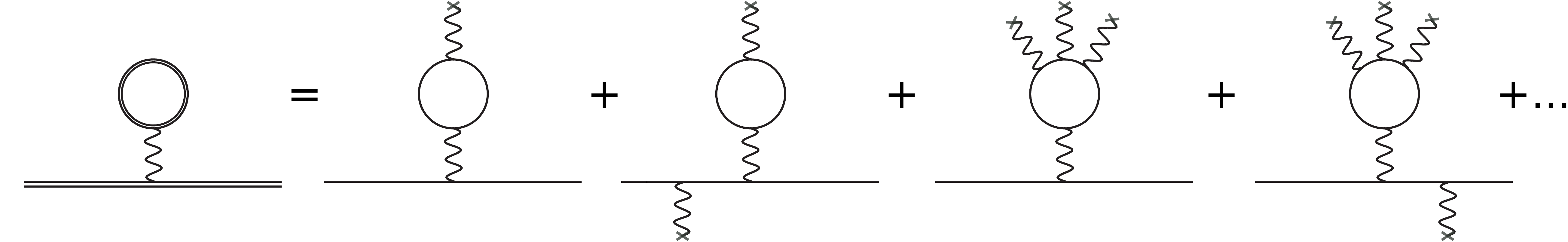}
	\caption{\label{fig-expansion} Diagrammatic representation of the weak-field expansion for the 1PR contribution to the electron propagator in a constant magnetic field. An odd number of external photons are attached to the loop (due to Furry's theorem) and an arbitrary number can be attached to the line.}

\end{figure}

The parameter integrals in (\ref{igen}) may be done numerically. As for the case of the dumbbell, however, it is instructive to expand in powers of a weak background field. Now we use $\cot z' -z'\csc^{2}z' = -\frac{2}{3}\left(z' + \frac{2}{15}z'^{3} + \ldots\right)$ which allows us to compute the $s'$ integral term by term. To cubic order in the magnetic field the $s'$ integral provides a factor
\begin{align}
	&\int_{0}^{\infty}\ud s^{\prime}  (4i \pi s^{\prime})^{-\frac{D}{2}} \e^{-im^{2}s^{\prime}}
\big(\cot z'-z'\csc^2 z'\big) \non
	&=\frac{2}{3 m^{2}} \left(\frac{ m^{2}}{4\pi}\right)^{\frac{D}{2}}\Big[\left(\frac{eB}{m^{2}}\right) \Gamma\Big[2 - \frac{D}{2}\Big] - \left(\frac{eB}{m^{2}}\right)^{3} \frac{2}{15} \Gamma\Big[4 - \frac{D}{2}\Big]  + \ldots \Big] \;.
\end{align}
The next step is to expand the $s$-integrand in $z$, the results of which we record in Appendix B. The remaining proper time integral over $s$ then yields
\begin{align}
	&S^{(1)1PR}(p)\approx\frac{2ie^{2}}{3} \left(\frac{m^{2}}{4 \pi}\right)^{\frac{D}{2}}\left[ \eBm \Gamma\Big[2 - \frac{D}{2}\Big] - \frac{2}{15} \eBm^{3}  \Gamma\Big[4 - \frac{D}{2}\Big]+ \ldots \right]\non
	&\times\bigg[ \frac{1}{4 m^{2}} \frac{ \big\{ m - \ps\,, [\gamma^{2}, \gamma^{1}] \big\}}{(m^{2} + p^{2})^{2} } +4i \left(\frac{eB}{m^{2}}\right) \left(  \frac{(m-\ps)p_{\perp}^{2}}{(m^{2} + p^{2})^{4}} -  \frac{p^{1}\gamma^{1} + p^{2}\gamma^{2}}{(m^{2} + p^{2})^{3}}\right)  + \ldots \bigg] \;,
	\label{S1B}
\end{align}
which is also represented diagrammatically in figure \ref{fig-expansion}. Here the top line is the contribution from the loop; the first term in square brackets diverges in $D = 4$. However, being linear in the coupling to the background this can be absorbed by a renormalisation. The first non-trivial contribution is of order $(eB/m^2)^3$, {which would be extremely interesting to compare to the weak field expansion of the one-particle-irreducible contribution to the propagator. Moreover, for strong fields it is important to check the relative size of these contributions in relation to  the Ritus mass shift.}

%%%%%%%%%%%%%%
\subsection{Scalar QED}
%%%%%%%%%%%%%%
The scalar QED expressions are slightly simpler. Using the projectors in (\ref{L2sc}) and after some straightforward manipulations detailed in the appendix, we arrive at the 1PR contribution to the two-loop EHL
\begin{align}
	\mathcal{L}^{(2)1PR} &= -\frac{2e^{2}}{D}\int_{0}^{\infty}\!\ud s(4\pi i s)^{-\frac{D}{2}}\e^{-i m^{2}s} \mathcal{J}_{\textrm{sc}}(z) \int_{0}^{\infty}\!\ud s'(4\pi i s')^{-\frac{D}{2}}\e^{-i m^{2}s'} \mathcal{J}_{\textrm{sc}}(z') \;,
	\label{L2scal}
\end{align}
where $\mathcal{J}_{\textrm{sc}}(z) = (z/\sin z)(\cot z - 1/z)$. As for the spinor case this involves physical contributions beyond renormalisation. To see this we give the leading contributions in an expansion in powers of the background field,
\begin{align}\label{igen2}
	\mathcal{L}^{(2)1PR} &= -\frac{2e^{2}}{Dm^{4}}\bigg(\frac{m^{2}}{4 \pi}\bigg)^D \bigg[\frac{1}{3}\eBm \Gamma\Big[2- \frac{D}{2}\Big] - \frac{7}{90}\eBm^{3}\Gamma\Big[4 - \frac{D}{2}\Big] + \cdots \bigg]^{2}\,,
\end{align}
where again each factor corresponds to one of the loops. Similarly, the first term in each can be subtracted by renormalisation, so that the physical contributions begin at order $(eB/m^2)^6$. These terms have been overlooked in previous work. The expanded result~(\ref{igen2}) may be represented by Feynman diagrams in the same way as for the spinor case, figure~\ref{figL2B}.

We may again analyse the strong field limit after the renormalisation $\mathcal{J}_{\textrm{sc}}(z) \rightarrow \mathcal{J}_{\textrm{sc}}(z) + \frac{z}{3}$ which renders the Euclidean space integral
\begin{equation}
	\frac{1}{eB} \left( \frac{4\pi}{eB} \right)^{-\frac{D}{2}} \int_{0}^{\infty} dz\, \e^{-\frac{m^{2}}{eB} z} z^{-\frac{D}{2}} \left[\textrm{csch}z - z\, \coth z + \frac{1}{3}z \right]
\end{equation}
finite in $D = 4$. It is not necessary to evaluate the integral since we know that the asymptotic behaviour can be extracted from the strong field limit of the finite (in $\epsilon$) part of the renormalisation term. Setting $D = 4-2\epsilon$ this takes the form
\begin{equation}
	\frac{1}{3 eB} \left(\frac{4\pi}{eB}\right)^{-2} \eBm^{\epsilon} \Gamma[\epsilon] + \textrm{ subleading},
\end{equation}
whose finite part is $\frac{eB}{48 \pi^{2}}\ln \eBm $ (the pole in $\epsilon$ is present only to cancel the original divergence of the integral (\ref{L2sc}) which, as in the spinor case, also provides some subleading contributions) so that the strong field behaviour is
\begin{equation}
	\mathcal{L}^{(2)1PR} \sim \frac{1}{2}B^{2}\left[\alpha \beta_{1} \ln \eBm\right]^{2}
\end{equation}
where now $\beta_{1} = \frac{1}{12\pi}$ is the first coefficient of the $\beta$-function in scalar QED. This is in agreement with \cite{Karbstein:2019wmj} and verifies that analysis to two-loop order. 

\begin{figure}
\centering
	\includegraphics[width=0.85\textwidth]{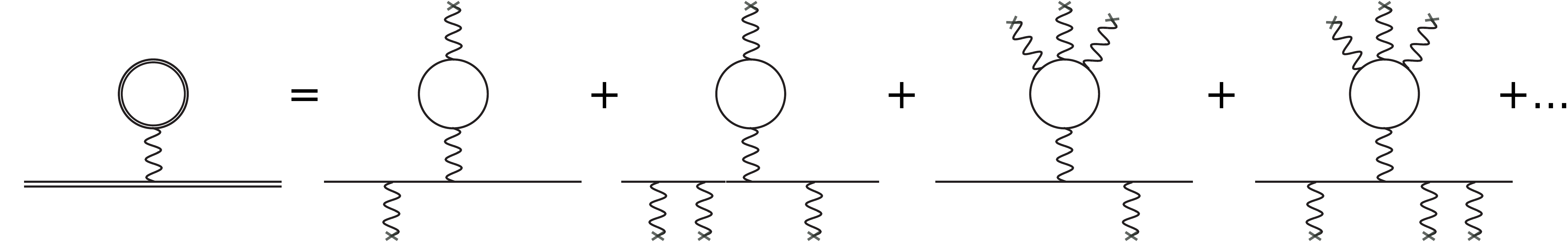}
	\caption{The weak field expansion of the reducible contribution to the scalar self-energy. A single photon attached to the loop can be renormalised away. Only an odd number of low energy photons couple to the line in Fock-Schwinger gauge.}
\label{fig-expansion-sc}
\end{figure}

Likewise, the 1PR contribution to the scalar propagator evaluates to
\begin{align}
	D_{\rm scal}^{1PR}(p) &= \frac{e^2}{2}\int_0^\infty \!\ud s'(4\pi is')^{-\frac{D}{2}}\e^{-im^2s'}\,\frac{z'}{\sin z'}\left({\rm cot} z'-\frac{1}{z'}\right) \label{D1sc-b} \\
\nonumber \times &\int_0^\infty \!\ud s\,s\, \e^{-im^2 s}\,\frac{1}{\cos z}\e^{-is(p_\parallel^2+\frac{\tan z}{z}p_\perp^2)}
 \bigg\{-\frac{s}{z}(\frac{{\tan}z}{z}-{\sec}^2z)p_\perp^2+i{\rm tan}z\bigg\}\,.	\nonumber
\end{align}
Now from here one can also take the weak field limit to obtain an expansion in powers of the coupling of the loop to the background,
\begin{align}
&D_{\rm scal}^{1PR}(p) = \frac{e^{2}}{2m^{2}} \left(\frac{m^{2}}{4\pi}\right)^{\frac{D}{2}}\Big[\frac{1}{3}\eBm \Gamma\Big[2-\frac{D}{2}\Big]-\frac{7}{90}\eBm^{3}\Gamma\Big[4 - \frac{D}{2}\Big] + \ldots \Big] \nonumber \\
&\times\int_0^\infty \!\ud s\, s\,\e^{-im^2 s}\,\frac{1}{\cos z}\e^{-is(p_\parallel^2+\frac{\tan z}{z}p_\perp^2)}\bigg\{-\frac{s}{z}(\frac{{\tan}z}{z}-{\sec}^2z)p_\perp^2+i{\rm tan}z\bigg\} \;,
\end{align}
which after performing the remaining proper-time integral yields	
\begin{align}
	&D_{\rm scal}^{1PR}(p) = \frac{e^{2}}{2} \left(\frac{m^{2}}{4\pi}\right)^{\frac{D}{2}}\Big[\frac{1}{3}\eBm \Gamma\Big[2-\frac{D}{2}\Big]-\frac{7}{90}\eBm^{3}\Gamma\Big[4 - \frac{D}{2}\Big] + \ldots \Big]  \nonumber \\
&\hspace{3cm}\times\Big[\eBm  \left(\frac{2p_\perp^2}{(m^2+p^2)^4}-\frac{1}{(m^2+p^2)^3}\right) \label{D1sc-lowb} \non
&\hspace{2em}+4m^{4}\eBm^{3}\left( \frac{5}{\left(m^{2} +p^{2}\right)^{5}} - \frac{36 p_{\perp}^{2}}{(m^{2} + p^{2})^{6} }+ \frac{40 p_{\perp}^{4}}{(m^{2} + p^{2})^{7}} \right)+\ldots \Big] \;.
\end{align}
See Appendix \ref{secApp1} for details of the computation and expansions. As before there is a piece linear in the coupling of the background to the loop that diverges in $D = 4$. This can be renormalised away. The remainder is a physical contribution to the scalar self-energy in a background magnetic field. To compare with the spinor result, we note that the second term in the second line of (\ref{S1B}), involving  $p_{\perp}^{2}/(m^2 + p^2)^{4}$ corresponds to the contribution of the 3-point scalar vertex to the spinor QED kernel that is also present in (\ref{D1sc-lowb}). Moreover, the powers of the coupling of the line to the background field (second set of square brackets) are now only odd, a reminder that the proper-time representation of the propagator was determined in Fock-Schwinger gauge. We show the expansion in figure \ref{fig-expansion-sc}.

% !TEX root = manuscript.tex
\section{1PR corrections in background plane waves}

\label{secPlane}
In this section we will investigate 1PR contributions in background plane waves of arbitrary strength and shape, which are used as models of intense laser fields. It is clear that for plane waves the (renormalised) Euler-Heisenberg effective action is zero (to any loop order, independent of whether it comes from 1PI or 1PR diagrams), because there are no Lorentz invariants which can be formed from the plane wave field strength alone~\cite{Schwinger,Dunne:2008kc}. The situation for the 1PR tadpole correction to a given diagram is less obvious; it is certainly possible to construct non-trivial invariants when there are other (momentum) vectors in play. Here we will use the worldline formalism to calculate the tadpole correction to, notably, \textit{any} process in a plane wave background. We will see that the tadpole gives a nonzero contribution, but that this can be renormalised away. Moreover this study provides an example of a non-constant background with a smooth limit to the crossed field case above.

We begin by defining the plane wave background. Given a lightlike direction $n_\mu$, $n^2=0$, we can always choose coordinates such that $n \cdot x  \equiv x^\LCp = x^0+x^3 $, and then the remaining coordinates are $x^\LCm := x^0-x^3$, ``longitudinal,'' and $x^\LCperp :=\{x^1,x^2\}$, ``transverse.''  A plane wave may be defined by a \textit{transverse} potential, $a_\mu(n\cdot x)$, so that $n\cdot a(n \cdot x)=0$, with field strength
\be
	eF_{\mu\nu} = n_\mu a'_\nu(n\cdot x) - a'_\mu(n \cdot x) n_\nu \;.
\ee
All plane waves obey
\be
	n^\mu F_{\mu\nu} = 0 \;, \qquad F_{\mu\nu}F^{\mu\nu} = F_{\mu\nu}{\tilde F}^{\mu\nu} = 0 \;. 
\ee
%
%Taking $a_\mu = \varepsilon_\mu n.x$ for constant $\varepsilon_\mu$ recovers the constant crossed field limit.

\subsection{The QED tadpole in a plane wave background}
%
%The worldline action is
%
%\bea
%	S &=& -m^2\frac{T}{2} - \int\limits_0^1\!\ud\tau\, \frac{\dot{x}^2(\tau)}{2T} + \dot{x}^\mu(\tau) a_\mu(x^\LCp(\tau)) \;, 
%\eea
%
The final expression for the tadpole correction to any diagram in plane wave backgrounds is simple, but to derive it using the worldline formalism requires a small departure from the methods commonly used for, and that are particular to, constant fields. We instead follow~\cite{Ilderton:2016qpj} which established a useful method of calculation for plane wave backgrounds.  First, we do not rotate to Euclidean space\footnote{Due to the arbitrary dependence of $a_\mu$ on $x^\LCp$ this would not give a positive definite action.}. Second, we do not use Fock-Schwinger gauge. (The choice of potential above makes the physics of particle dynamics in the wave manifest, see~\cite{Dinu:2012tj}.)  Third, the worldline Green function in the plane wave background will not be needed. Instead we will perform the required coordinate-space integrals defining the tadpole contribution directly, using a suitable basis of functions on the unit circle. In this section the dimensional regulation of the proper-time integrals is left implicit; it can be made explicit by analytically continuing in the number of \textit{transverse} directions, which preserves the tensor structure of the plane wave, see~\cite{Casher:1976ae,Brodsky:1997de} for details and~\cite{Ilderton:2013dba} for an application in plane wave backgrounds.

From~\cite{GKUs1,GKUs2} the tadpole part, Fig.~\ref{figTad}, of any QED Feynman diagram in the presence of a plane wave background may be  written
\be\label{amp1}
	\Gamma_1 = -2 \int\limits_0^\infty\frac{\ud s}{s}\Anton{\oint\mathcal{D}^4x}\exp\bigg[-im^2s-i\int\limits_0^1\!\ud\tau\ \frac{\dot{x}^2}{4s} + \mathcal{A}\cdot \dot{x}\bigg]\, \text{Spin} \bigg|_{\mathcal{O}(\epsilon)}\;, 
\ee
where $m$ is the mass of the particle in the tadpole loop and $\mathcal{A}_\mu=a_\mu(x^\LCp)+e \epsilon_\mu e^{-ik \cdot x}$, with $k^{\mu}$ and $\epsilon_{\mu}$ the momentum and polariastion of the attached photon and $e$ the electromagnetic charge. $\text{Spin}$ is shorthand for the Feynman spin factor~\cite{Feynman:1951gn,ChrisRev} that couples the spin degrees of freedom to the electromagnetic field,
\be\label{Spin}
	\text{Spin}=\frac{1}{4}\text{tr}_{\gamma}\mathcal{P}\exp\bigg[-\frac{is}{2}\int\limits_0^1\ud\tau\, \sigma^{\mu\nu}\mathcal{F}_{\mu\nu} \bigg]\;,
\ee
with $\mathcal{F}$ the field strength derived from $\mathcal{A}$. The trace is over the Dirac matrices ($\sigma^{\mu\nu}=\frac{i}{2}[\gamma^\mu,\gamma^\nu]$ are the spinor generators of the Lorentz group), and $\mathcal{P}$ stands for path-ordering. The functional integral is over closed trajectories in Minkowski space, $x(\tau)$, on the unit circle $x^\mu(0)=x^\mu(1)$. The variable $s$ parameterises the invariant length of the worldline (the Schwinger proper time) and is also to be integrated over. The prescription indicated by $\big|\raisebox{-0.25em}{$\mathcal{O}(\epsilon)$}$ is that one takes only the piece that is linear in the photon polarisation vector $\epsilon$. We break the calculation of $\Gamma_1$ into the following stages, before attaching it to an (arbitrary) Feynman diagram in Sect~\ref{sew-sect}.

\subsubsection*{Simplifying the spin factor}

Although it is common to employ a Gaussian (Grassmann) integral representation of the spin factor~\cite{ChrisRev}, it is simpler and more direct here to use the representation above, as many simplifications will follow from the plane wave structure, see e.g.~\cite{Dunne:2005sx,Ilderton:2016qpj}.

There are two terms linear in $\epsilon$ in (\ref{amp1}). We can take an $\epsilon$ from $\mathcal{A}$ in the exponential or we take an $\epsilon$ from $\mathcal{F}$ in Spin. So to proceed we expand the Spin and write down the possible terms using the explicit result
\be\label{explicit}
	-\frac{i}{2}\sigma^{\mu\nu}\mathcal{F}_{\mu\nu}=\slashed{n}\slashed{a}'(x)-ie(\slashed{k}\slashed{\epsilon} - k \cdot \epsilon )\e^{-ik \cdot x} \;.% \equiv  \mathcal{F}_0(\tau) + \mathcal{F}_\epsilon(\tau) \;.
\ee
Consider the $N^\text{th}$ order term in the expansion of the exponential of (\ref{Spin}), containing $N$ powers of (\ref{explicit}), from which we wish to extract the terms up to $\mathcal{O}(\epsilon)$. As we show in Appendix C, we only need retain the $N=0$ and $N=2$ terms which may be evaluated directly. A convenient form of the resulting contributions is
%\J{I notice that for the unit circle, one could actually pull the parameter integral outside of the brackets}
%
\be\label{gamma-red-1}
	\Gamma_1 = 2e\int\limits_0^\infty\frac{\ud s}{s}\oint\mathcal{D}^4x \! \int\limits_0^1\!\ud\sigma \, \e^{iS-i \int J \cdot  x \, d\tau } \bigg[  1 + ie s^2 \int\limits_0^1\!\ud\tau\, \epsilon \cdot F(x(\tau)) \cdot k  \bigg]\bigg|_{\mathcal{O}(\epsilon)} \;,
\ee
in which, %following an integration by parts,
\be
	S = -m^2s - \int\limits_0^1\!\ud\tau\,\bigg[ \frac{\dot{x}^2}{4s} + \dot{x} \cdot a(x^\LCp)\bigg]; \quad
	J_\mu(\tau) = k_\mu \delta(\tau-\sigma) + \epsilon_\mu \dot{\delta}(\tau-\sigma) \;. \label{SJ}
\ee
The representation (\ref{gamma-red-1}) makes clear the relative contribution of spin effects, because if we delete the second term in square brackets, we obtain the sQED expression (up to an overall constant). See also below.

\subsubsection*{Coordinate integrals}
To carry out the path integration over the closed trajectories we split the coordinates into a centre of mass piece $x^\mu_c$ and an orthogonal fluctuation $y^\mu$, which in particular helps deal with the zero mode of the functional integral -- see the appendix. Just as in \cite{Ilderton:2013dba}, performing the $x_c^\LCperp$ and $x^\LCm_c$ integrals produces a delta function fixing the photon momentum to lie in the laser momentum direction\footnote{Covariant indices are $p_\LCpm = (p_0\pm p_3)/2$ and $p_\LCperp = \{p_1,p_2\}$. Measures obey $\ud^4 x^\mu = \ud x^{\LCp} \ud x^\LCm \ud^2 x^\LCperp /2$ and $\ud^4 p_\mu = 2\ud p_{\LCp} \ud p_\LCm \ud^2 p_\LCperp$.}, $k_\mu = n_\mu k_\LCp$ (see (\ref{varje}) in the appendix for normalisation conventions);
\be
	\oint\!\mathcal{D}^4x \ldots = (2s)^2 \int\!\ud^4 x_c \oint\!\mathcal{D}^4 y \ldots = \frac{1}{2}(2\pi)^3\delta^3_{\LCm,\LCperp}(k)(2s)^2 \int\!\ud x^\LCp_c \oint\!\mathcal{D}^4 y \ldots
\ee
However, we may \textit{not} yet use the delta function to simplify expressions, because of the singular structure in the sewing integral which attaches the tadpole to a larger diagram. We next perform the $y^\LCm$ integral. To do so we first shift variables in $y^\LCp$, writing $y^\LCp=y^\LCp_\text{cl} + \delta y^\LCp$ where $y^\LCp_\text{cl}$ is the classical path obeying the equations of motion
\be
	\ddot{y}^\LCp_\text{cl} = 4 s J_\LCm = 2s J^\LCp \implies y^\LCp_\text{cl}(\tau) = 2s \int\limits_0^1\ud\tau'\, G_{\tau\tau'}J^\LCp(\tau') \;,
\ee
with $G_{\tau\tau'}$ the \textit{free} worldline propagator on the space of fluctuations that is given in the appendix. Since the solution $y^\LCp_\text{cl}$ always appears together with $x_c^\LCp$ we define
\be
\begin{split}
	\varphi(\tau) :&= x_c^\LCp + y^\LCp_\text{cl}(\tau) \\
	&=x^\LCp_c + 2n \cdot k\, s G_{\tau\sigma} + 2n \cdot \epsilon\, s \partial_\tau G_{\tau\sigma} \equiv \varphi_0(\tau) + \varphi_1(\tau) \;,
\end{split}
\ee
in which the subscripts refer to the order in $\epsilon$ of the terms. Following this shift the only $y^\LCm$ dependence in $\Gamma_1$ appears in the exponent as
\be\label{hej-delta}
	i \int\limits_0^1\!\ud\tau\, y^\LCm \frac{\delta \ddot{y}^\LCp}{4s} \;.
\ee
The integral over $y^\LCm$ produces a delta functional that sets, because of the periodic boundary conditions, $\delta y^\LCp=0$. (The same is seen in the calculation of helicity flip in a plane wave~\cite{Ilderton:2016qpj}. For related simplifications in pair production see~\cite{Ilderton:2014mla}, and also~\cite{Halpern:1976gd,Halpern:1977he}.) In order to keep track of factors of $s$, it is simplest to leave the integral over (\ref{hej-delta}) unevaluated, for now, and to set $\delta y^\LCp \to 0$ in the rest of the amplitude. 

From here we adopt the following notation for averages on the unit circle:
\be
	\int\limits_0^1\!\ud \tau f(\tau) \to \langle f \rangle \;,
	\quad 
	\int\limits_0^1\!\ud \tau' G_{\tau\tau'}f(\tau') \to \langle G_{\tau \bullet} f \rangle \;,
	\quad 
	\int\limits_0^1\!\ud \tau\ud \tau' g(\tau) G_{\tau\tau'}f(\tau') \to \langle g, Gf \rangle \;.
\ee
We now similarly shift the perpendicular coordinates by the classical solution obeying
\be
	{\ddot y}_\text{cl}^\LCperp=2s(J^\LCperp - {\dot a}^\LCperp) \implies y_\text{cl}^\LCperp(\tau) = 2s \langle G_{\tau\bullet}\big(J^\LCperp - {\dot a}^\LCperp(\varphi)\big)\rangle \;.
\ee
The effect of this shift is to collect all dependence of the fluctuations $\delta y$ into
\be\begin{split}
%\oint\!\mathcal{D}^4 \delta y\, \exp\bigg[-\frac{i}{2T}\int\limits_0^1\!\ud\tau\ \delta\dot{y}^2\bigg] \;,
\oint\!\mathcal{D}^4 \delta y\, \exp\bigg[-\frac{i}{4s}\langle \delta\dot{y}^2\rangle \bigg] = {-(2\pi)^{-2} (2s)^{-4}}\;,
\end{split}
\ee
in which the Gaussian integral is that of the free theory, see (\ref{varje})--(\ref{pathnorm}) in the appendix. At this stage we have obtained
\be\label{mer}
	\Gamma_1 = {-}\frac{\pi e}{2} \delta^3_{\LCm,\LCperp}(k)\int\limits_0^\infty\frac{\ud s}{s^3}\int\!\ud x^\LCp_c \! \int\limits_0^1\!\ud\sigma \, \e^{iW} \bigg[  1 + i es^2 \int\limits_0^1\!\ud\tau\, \epsilon\cdot F(\varphi(\tau)) \cdot k  \bigg]\bigg|_{\mathcal{O}(\epsilon)} \;,
\ee%
where what remains in the exponent,  $W$, is defined by
\be
	W = -s \big\langle J_\mu - \dot{a}_\mu(\varphi) , G(J^\mu - \dot{a}^\mu(\varphi)) \big\rangle\;.
\ee
It can be checked that $W$ is the \textit{classical} action. That the functional integrals lead to the classical action is due to the many symmetries of the background.

\subsubsection*{Expansion to order $\epsilon$}
We now expand the exponential of (\ref{mer}) to order $\epsilon$. We begin with (writing $\delta_{\tau \sigma} := \delta(\tau - \sigma)$ for brevity)
\be\begin{split}
	J_\mu - \dot{a}_\mu &=  {{k_\mu \delta_{\tau\sigma} -\dot{a}_\mu(\varphi_0(\tau))}}+ {{\partial_\tau \bigg( \epsilon_\mu \delta_{\tau\sigma} - \frac{n \cdot \epsilon}{n \cdot k}\dot{a}_\mu(\varphi_0(\tau))\bigg)}} + \mathcal{O}(\epsilon^2) \\
	&\equiv \alpha_\mu + \beta_\mu + \mathcal{O}(\epsilon^2)\;,
\end{split}
\ee
where $\alpha_\mu$ ($\beta_\mu$) is order zero (one) in $\epsilon$. Thus to order $\epsilon$ we have
\be\label{W-expansion}
	%e^{iW} \to \big\{-i T \langle \alpha_\mu, G\beta^\mu \rangle , 1 \big\} \, \exp\bigg[-\frac{iT}{2} \langle \alpha_\mu, G \alpha^\mu\rangle \bigg] \;,
	\e^{iW} \to -2i s \langle \alpha_\mu, G\beta^\mu \rangle \exp\big[-is \langle \alpha_\mu, G \alpha^\mu\rangle \big] \;,
\ee
in the \textit{first} term in large square brackets of (\ref{gamma-red-1}), while for the second term in large square brackets, which is already linear in $\epsilon_\mu$, we replace everything outside the exponential in (\ref{W-expansion}) with unity. The exponential terms are, using periodicity,
\be\begin{split}
	\langle \alpha_\mu, G \alpha^\mu\rangle &= k^2 G_{\sigma\sigma} + \langle a_\mu(\varphi_0)\rangle \langle a^\mu(\varphi_0)\rangle - \langle a_\mu(\varphi_0) a^\mu(\varphi_0) \rangle \;. \\
		&\equiv -\frac{1}{12}k^2 - \text{var}(a) \;.
\end{split}
\ee
The $a_\mu$-dependent terms are a variance, generating the ``effective mass'' of a particle in a plane wave background~\cite{Kibble:1975vz,Harvey:2012ie,Ilderton:2016qpj}. The factor of $-1/12$ is the coincidence limit $G_{\sigma\sigma}$; such contributions are usually assumed to be killed in vacuum by the overall momentum conserving delta function. Indeed note that all terms vanish if we use this delta function, for then $k^2 = n \cdot k = 0$ from the start. Again, though, we may not use such arguments until we have sewn the tadpole onto a larger diagram, as otherwise we risk missing precisely the 1PR contributions of interest.

We turn to the pre-exponential factor in (\ref{W-expansion}). Integrating by parts, using periodicity of $\varphi_j$, and that $\dot{G}_{\sigma\sigma}=0$, % (which kills a free-field contribution),
we find\footnote{In simplifying the pre-exponential terms one encounters the $\tau$-integral of $a^\mu(\varphi_0(\tau)) \partial_{\tau} \varphi_0(\tau)$, which is exact and therefore vanishes by the periodicity of $\varphi_0$.}

\be\label{pre-exp}
	\langle \alpha_\mu, G \beta^\mu\rangle  = \bigg(\epsilon_\mu - \frac{n\cdot \epsilon}{n \cdot k} k_\mu \bigg) \big(a^\mu(\varphi_0(\sigma)) - \langle a^\mu(\varphi_0)\rangle \big) \;,
\ee
which is linear in $a_\mu$ and, it can be checked, independent of $\sigma$. 
%
%%%%%%%
\subsection*{Final result and simplification}
%%%%%%%
%
At this stage we can add (\ref{pre-exp}) to the piece from the spin factor in (\ref{gamma-red-1}) that is already linear in the polarisation vector. As the spin factor only depends on $x^\LCp(\tau)$, the preceding calculation of the path integral goes through without change, and due to (\ref{hej-delta}) we simply replace $x^\LCp(\tau) \to \varphi_0$ in the spin factor (since this part is already linear in $\epsilon$ we drop the $\varphi_{1}$ part). At this stage we can write out the full, but cumbersome expression for the tadpole, from here on dropping the subscript ``$c$'' on the centre of mass piece $x^\LCp_c \to x^\LCp$,
\be\label{gamma-f2}
\begin{split}
&\Gamma_1= -2ie\pi\, \delta^3_{\LCperp,\LCm}(k)
\int\limits_0^\infty \frac{\ud s}{s} \e^{-i s m^2}
\int\!\ud x^\LCp \e^{-ik_\LCp x^\LCp} \e^{is\,\text{var}(a)} \e^{is\tfrac{k^{2}}{12}} \\
&\times \bigg[-\frac{1}{2s}\bigg(\epsilon_\mu - \frac{n\cdot \epsilon}{n\cdot k} k_\mu \bigg)\,\!\! \big(a^\mu(\varphi_0(\sigma)) - \langle a^\mu(\varphi_0)\rangle\big)  + \frac{e}{4} \epsilon \cdot \langle F(\varphi_0)\rangle \cdot k \bigg] \;.
\end{split}
\ee
This is a non-trivial function of the background field, containing arbitrary powers of the gauge potential due to the exponent. However, the only relevant part of $\Gamma_1$ is that which survives being sewn to another diagram. By considering the dependence of the various parts of (\ref{gamma-f2}) on $k_\mu$ we will shortly find a considerably simpler expression for this surviving contribution. We have the following properties.
\bi
	\item[P1.] Expanding the field-dependent exponential involving $\textrm{var}(a)$ generates (one plus) $x^\LCp$- and $s$-dependent terms with $n\geq 2$ powers of $a_\mu$. These terms could contribute physical (i.e.~non-renormalisation) effects to other diagrams. It can be checked directly that each such term comes with at least \textit{two} powers of $n\!\cdot\! k$.
	\item[P2.] Expanding the pre-exponential terms, i.e.~the second line in (\ref{gamma-f2}), generates leading order contributions proportional to $\epsilon \!\cdot\! F(x^\LCp) \!\cdot\! k$, and then $x^\LCp$- and $s$-dependent terms containing higher $x^\LCp$-derivatives of $\epsilon \!\cdot\! F(x^\LCp) \!\cdot\! k$. Each derivative comes with an additional power of $\varphi_0$ and, thought this, a power of $n \!\cdot\! k$. 
	\item[P3.] Expanding the final exponent in (\ref{gamma-f2}) contributes (one plus) powers of $k^2$.
\ei
Making the expansions above, we may write $\Gamma_1$ as
\be\begin{split}	\label{G1Pre}
	\Gamma_1 = {\frac{1}{3}} ie^2 \pi \delta^3_{\LCperp,\LCm}(k)\int \limits_0^\infty\! &\frac{\ud s}{s}\,\e^{-ism^2} \!\!\int\!\ud x^\LCp \e^{-ik_\LCp x^\LCp} k \!\cdot\! F(x^\LCp) \!\cdot\! \epsilon  \\
	&+ \text{higher powers in $a$} + \text{derivative terms} + k^2 \text{ terms.}
\end{split}
\ee
We now perform the $x^\LCp$ integral. For the term shown this gives the Fourier transform $\tilde{F}_{\mu\nu}(k_\LCp)$. The higher order terms have a more complicated functional dependence, but nevertheless are just Fourier transforms. Note that the essential tensor structure of all the terms is given by derivatives of $k \cdot F \cdot \epsilon$. It is convenient to introduce an auxiliary variable $\omega$ and write the Fourier transform as an integral over a delta function setting $\omega \to k_\LCp$, in order to obtain a covariant $\delta^4$, thus:
\be
	\delta^3_{\LCperp,\LCm}(k)   f(k_\LCp)  = \delta^3_{\LCperp,\LCm}(k)   \int\! \ud \omega \, f(\omega) \delta(k_\LCp - \omega)  = 2 \int\! \ud \omega \,  f(\omega) \delta^4(k - l)   \;,
\ee
in which $l_\mu := \omega n_\mu$, here and below, is an auxiliary momentum. It follows that $\Gamma_1$ has the expansion
\be\label{GSPINKLAR}
	\Gamma_1 = {\frac{2}{3}} ie^2 \pi \int \limits_0^\infty \frac{\ud s}{s}\,\e^{-ism^2}\!\!\int\!\ud \omega\, \delta^4(k-l) \, k \!\cdot\! {\tilde F}(\omega) \!\cdot\! \epsilon + \ldots \;,
\ee
in which the ellipses denote the ``higher order'' terms summarised in (\ref{G1Pre}). Note that if, at any stage of this calculation, we had taken the delta functions on trust, then we would have obtained zero for the tadpole. However, we must first sew the tadpole to another diagram.

\subsection{The tadpole correction to any diagram}\label{sew-sect}
\begin{figure}[t!]
	\centering\includegraphics[width=0.3\textwidth]{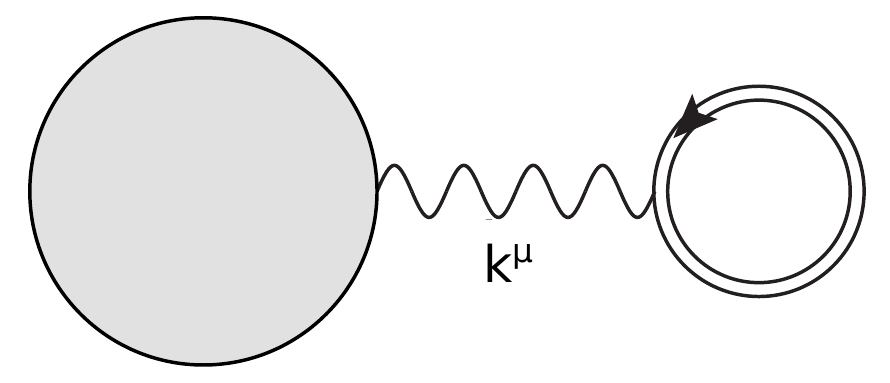}
	\caption{\label{FIG:ALL} The one-loop tadpole correction to any diagram.}
\end{figure}
We now show that none of the ``higher order'' terms neglected in (\ref{G1Pre}) or (\ref{GSPINKLAR}) can survive being sewn. Consider the tadpole correction to any diagram, as illustrated in Fig.~\ref{FIG:ALL} (one could keep in mind sewing the tadpole to a particle propagator as we have done above, for which the one-particle \textit{irreducible} contributions have previously been calculated~\cite{Meuren:2011hv,Ilderton:2013tb}.) The photon with momentum and polarisation $k_\mu$ which is part of our tadpole is attached at its other end to the larger diagram. We write the tadpole as $\Gamma_1= \Gamma_1^\mu(k) \epsilon_\mu$, and similarly write the rest of the diagram as $\Delta^\mu(-k) \epsilon_\mu$. This is also linear in the photon polarisation by the assumption that the tadpole's photon is attached to it. Then the sewing prescription in Feynman gauge is to make the replacement $\epsilon_{\mu} \epsilon_{\nu} \rightarrow \eta_{\mu\nu}/ {k^2}$ and then to integrate over the intermediate photon momentum $k_\mu$. The 1PR contribution to the two-loop effective action is then
\be
	\Gamma^{(2)1PR}_{\textrm{spin}} =	\int\!\frac{\ud^4 k}{(2\pi)^4 k^2} \Delta_\mu(-k) \Gamma_1^\mu(k) \;.
\ee
We can see from (\ref{GSPINKLAR}) that the photon connecting the tadpole to another diagram has support not just at $k_\mu=0$ as for constant fields, but rather at $k_\mu = l_\mu = \omega n_\mu$. However, since $l^2=0$, on-shell, the support of the delta function is still precisely where the $k^2$ denominator in the sewing integral vanishes. We require, then, a generalisation of the sewing relation (\ref{SEW-1}) which allows us to extract the finite part of the singular sewing structure, beyond the case of constant fields.

Now, the diagram to which we sew will in general be a function of $k_\mu$ multiplied by, because the process occurs in a plane wave, a three dimensional delta function $\delta^3_{\LCperp,\LCm}(P-k)$ where $P$ is some collection of momenta. This can always be made covariant as for the tadpole, above. It is safe to use the delta function coming from the tadpole to either replace $\delta^3_{\LCperp,\LCm}(P-k) \to \delta^3_{\LCperp,\LCm}(P)$, or to replace a covariant $\delta^4(P-k) \to \delta^4(P-\omega n)$ and to bring this inside the $\ud \omega$ integral in (\ref{GSPINKLAR}), as this does not affect the singular structure in the sewing integral. Hence, the type of sewing integral we encounter is, for $l^2=0$, 
\be\label{1}
\begin{split}
\mathcal{K}_{\mu_1 \mu_2 \cdots \mu_n} &:= \int\!\ud^4 k\, \frac{1}{k^2} \, \delta^4(k - l) \, k_{\mu_1} k_{\mu_2} \cdots k_{\mu_n} \\
&= c_n \big( g_{\mu_1\mu_2} l_{\mu_3} \cdots l_{\mu_n}  + \;  \text{symmetrised}\big) + \lambda_n l_{\mu_1} l_{\mu_2} \cdots  l_{\mu_n} \;,
\end{split}
\ee 
where the constant $c_n$ is determined by taking the trace, while the constant $\lambda_n$ will always drop out (the equality of these expressions is shown in the appendix). The most important case is
\be
	\mathcal{K}_{\mu\nu} =  \frac{1}{4}g_{\mu\nu} + \lambda_2 l_\mu l_\nu \;.
\ee 
We have the following results
\be\begin{split}\label{stick-resultat}
	&\text{i) } n^\mu n^\nu \mathcal{K}_{\mu\nu \cdots \tau} = 0 \;, \qquad \quad \, \text{ii) } \mathcal{K}^\alpha_{\alpha \ldots \tau} F^{\tau\rho} =0 \;, \\
	&\text{iii) } n^\mu \mathcal{K}_{\mu\nu \cdots \tau} F^{\tau\rho} =0 \;, 
	\qquad \text{iv) }\mathcal{K}^\alpha_{\alpha \ldots \tau} n^\tau =0 \;.
	% \mathcal{K}^\alpha_{\alpha \ldots \tau} \not=0, 
\end{split}
\ee
Given this, consider again the expansion of (\ref{gamma-f2}) into (\ref{GSPINKLAR}) plus corrections, and the sewing to a larger diagram. Any term in the tadpole containing $(n\cdot k)^2$, or higher powers thereof, will vanish when sewn, by i) of (\ref{stick-resultat}). Hence, from P1, no higher power of the field strength can survive sewing. All derivative terms from the second line of (\ref{gamma-f2}) at greater than linear order vanish for the same reason. The \textit{first} derivative term vanishes because of iii).  Hence, from P2, no derivative terms in the tadpole survive sewing. Finally, no term containing $k^2$ can survive because of ii) and iv), see P3, above. It follows that the only part of the tadpole which survives being sewn to another diagram is (\ref{G1Pre}), equivalently (\ref{GSPINKLAR}).

Hence, let any diagram have the ``$k$-linear'' part $\delta^4(P-k)\Delta(P)_{\mu\nu}k^\mu \epsilon^\nu$. Then the tadpole correction is
\begin{align}
	\Gamma_{\textrm{spin}}^{(2)1PR} &= {\frac{2}{3}} ie^2\pi\!\int \limits_0^\infty \frac{\ud s}{s}\,\e^{-ism^2}\!\!\int\!\ud \omega\,   \delta^4(P-\omega n) \Delta_{\mu\nu}(P) {\tilde F}(\omega)_{\sigma\nu} \!\int\!\frac{\ud^4 k}{(2\pi)^4k^2}  \delta^4(k-l) \, k^\mu  k^\sigma \nonumber \\
	&= {\frac{1}{6}} ie^2\pi\, 	\int \limits_0^\infty \frac{\ud s}{s}\,\e^{-ism^2}\int\! \frac{\ud \omega}{(2\pi)^4}  \delta^4(P-\omega n) \Delta_{\mu\nu}(P) {\tilde F}(\omega)^{\mu\nu} \;.
	\label{tadAny}
\end{align}
Crucially this is, as for the case of constant crossed fields, linear in the external field coupling to the tadpole loop, so it can be absorbed into a renormalisation. 

The most important aspect of the sewing, then, is that all terms of higher order in the background field vanish, so that although the tadpole itself involves the gauge potential to all order, the contribution that sees the larger diagram through the mediating photon is at most linear in the external field. It is not entirely obvious from the beginning that this should be the case, and indeed one can imagine other terms that could have contributed to the final result (\ref{tadAny}). For example, if $\tilde{F}^2$ (suitably normalised by powers of $m$ or $s$) had appeared in (\ref{tadAny}) then we would have non-renormalisation effects. Similarly, if the sewing allowed terms like $p\cdot F^2\cdot p$ for $p_\mu$ some momentum from the larger diagram, then we could have had arbitrary powers of the field strength in the final expression. However, we have seen that no such terms arise. Thus, we have found that although the tadpole contribution is nonvanishing in a plane wave background, it does not induce a physical correction to any process.
\subsection{The dumbbell and the effective Lagrangian}
Our expression for the tadpole allows us to examine the two-loop dumbbell diagram in plane waves, see Fig.~\ref{figDumb}. This could give a nontrivial contribution to the vacuum persistence amplitude~\cite{Brown:1964zzb} if the diagram developed an imaginary part -- it is well known, though, that there is no pair production in plane waves~\cite{Schwinger}. %which shall be verified in the following calculation.

We take two copies of (\ref{GSPINKLAR}), send $k_\mu \to -k_\mu$ in one, and sew them together:
\be
	\Gamma_{\textrm{spin}}^{(2)1PR}:=\int\!\frac{\ud^4 k}{(2\pi)^{4} k^2} \Gamma_1^\mu(-k)\Gamma_{1\,\mu}(k) \;.
\ee
Define the (dimensionally regulated) constant $c$ by
\be
	c = {\frac{2}{3}}ie^2\pi \int \limits_0^\infty \frac{\ud s}{s}\,\e^{-ism^2} \;.
\ee
Then we find for the dumbbell diagram
\be\begin{split}\label{GGtrams}
	\Gamma_{\textrm{spin}}^{(2)1PR} &= c^2 \int\!\frac{\ud^4k}{(2\pi)^4 k^2} \int\!\ud\omega \ud\nu\, \delta^4(k-\omega n)\delta^4(k+\nu n)\, k \cdot  \tilde{F}(\omega) \cdot \tilde{F}(\nu) \cdot k \\
	&= c^2
		 \int\!\ud\omega \ud\nu\,  \delta^4(\omega n +\nu n)
	\int\!\frac{\ud^4k}{(2\pi)^4 k^2}  \delta^4(k-\omega n) k \cdot \tilde{F}(\omega) \cdot \tilde{F}(\nu) \cdot k \\
	&=
		 \frac{c^2}{2}V^\LCperp V^\LCm \int\!\ud\omega
	\int\!\frac{\ud^4k}{(2\pi)^4 k^2}  \delta^4(k-\omega n)\, k \cdot  \tilde{F}(\omega) \cdot \tilde{F}(-\omega) \cdot k \;.
\end{split}
\ee
The volume of the longitudinal and transverse directions reflects the translation invariance of the process in those three directions. Interestingly, the integrand of (\ref{GGtrams}) is proportional to the (Fourier transformed) ``$\chi$-factor'' of the intermediate photon~\cite{RitusRev},
\be\label{chi}
 	\chi(\phi) = e^2k^\mu F_{\mu\sigma}(\phi)F^{\sigma}_{\,\,\,\nu}(\phi)k^\nu = -(n\cdot k)^2 a^{\prime 2}(\phi) \;,
\ee
which determines the relevance of nonlinear quantum effects in plane wave backgrounds~\cite{RitusRev,DiPiazza:2011tq,Seipt:2017ckc}. However, the whole expression is ultimately killed because the $k_\mu$ integral gives $\mathcal{K}_{\mu\nu} \sim g_{\mu\nu}$, above, which replaces $\chi \to \text{tr } (F^2) =0$, and the dumbbell vanishes.

This is reassuring since, if the dumbbell did not vanish, there could be a non-zero contribution to the Euler-Heisenberg effective Lagrangian for plane waves. To confirm that there is no such contribution, we take two copies of the tadpole, and reintroduce the centre of mass coordinates in each. We then define an average of these two positions, call it $x_0^\mu$, and extract the contribution to Euler-Heisenberg via the definition 
\be
	\Gamma_{\textrm{spin}}^{(2)1PR} := \int\!\ud^4x_0\, \mathcal{L}_{\textrm{spin}}(x_0) \;.
\ee
A straightforward extension of the dumbbell calculation yields
\be
	\mathcal{L}_{\textrm{spin}}(x_0) = \frac{c^2}{(2\pi)^4}\!\int\!\ud\omega\ud\nu\, \e^{i(\omega+\nu)n \cdot x_0} \!\!\int\!\frac{\ud^4k}{(2\pi)^4 k^2} \delta^4\big(k+\tfrac{1}{2}(\nu - \omega) n\big) k \cdot {\tilde F}(\omega) \cdot {\tilde F}(\nu) \cdot k \;,
	%\hspace{-4em} &\J{\sim \frac{c^2}{(2\pi)^4}\int\!\ud\omega\ud\nu\, e^{i(\omega+\nu)n.x_0} \int\!\frac{\ud^4k}{(2\pi)^4 k^2} \delta^4\big(k- \omega n\big) k \cdot {\tilde F}(\omega)\cdot {\tilde F}(\nu) \cdot k ,}
\ee
which again vanishes after performing the sewing integral.
%In this way one can adapt (\ref{tadAny}) to deduce $\Gamma_{\textrm{scal}}^{(2)1PR}$ and in turn compute the dumbbell for scalar matter. The result is simply $\mathcal{L}_{\textrm{scal}} = \frac{1}{4}\mathcal{L}_{\textrm{spin}}$..

\subsection{Examples: the constant crossed field limit and scalar QED}
The methods used here for the plane wave calculation are quite different to those used for constant fields, above. A mutual check on these methods is thus provided by re-deriving crossed field results from the general plane wave result. The crossed field is defined by $a_\mu(x^\LCp) = \varepsilon_\mu x^\LCp$ for $\varepsilon_\mu$ a spacelike constant (not to be confused with the polarisation vector of the attached photon). The constant field strength is then $eF_{\mu\nu} = n_\mu \varepsilon_\nu - \varepsilon_\mu n_\nu$. The variance in this case is
\be\begin{split}
	\text{var}(a) =  \frac{1}{180} s^2(n\cdot k)^2  \varepsilon_\mu \varepsilon^\mu  \;,
\end{split}
\ee
which is independent of $x^\LCp$. The second line of (\ref{gamma-f2}) simplies exactly to $-\tfrac{2}{3} k\cdot F \cdot \epsilon$, constant. Carrying out the $x^\LCp$ integral gives ($2\pi$ times) a fourth delta function, which is just the Fourier transform of the field, consistent with (\ref{GSPINKLAR}) and
\be\label{gamma-f3}
	\Gamma_1= \frac{{4i}e^2\pi^2}{3}\, \delta^4(k)\, k \cdot F \cdot \epsilon \;  \int\limits_0^\infty \frac{\ud s}{s}\, \e^{-is m^2 +i \frac{s^3}{180} (n\cdot k)^2\varepsilon^2 +i\tfrac{s}{12}k^2} \;.
\ee
Consider now sewing this onto a larger diagram, denote it by $\Delta$. The more general sewing integral above reduces to (\ref{SEW-1}). The only term which can survive this sewing is quadratic in $k_\mu$. Hence if $\Delta$ contributes a linear term, so $\Delta \supset  \epsilon_{\mu}\Delta^{\mu\nu} k_\nu$, then the tadpole can couple to this (any part independent of the momentum is killed by symmetry when integrated). Observe that we may therefore, without losing any terms, simplify the tadpole to
\be\label{gamma-f4}
	\Gamma_1 \to \frac{{4i e^2}\pi^2}{3}\, \delta^4(k)\, k \cdot F \cdot \epsilon \;  \int\limits_0^\infty \frac{\ud s}{s}\, \e^{-ism^2} \;,
\ee
exactly as argued for the general result (\ref{GSPINKLAR}). Moreover, were we to sew two tadpoles together to form the two-loop reducible contribution to the EHL of figure \ref{figDumb} then the momentum integral produces the contraction $\tr(F^{2})$, which vanishes, as for the general case (see text below (\ref{chi})). The same result follows if the crossed field limit is taken directly in (\ref{GGtrams}). This also reproduces the results of Sect.~\ref{secCrossed} as a smooth limit of a more realistic spatially varying field configuration. 

Finally, we comment on the scalar QED tadpole. Note that when identifying the the surviving contributions in $\Gamma_1$, we expanded the second line of (\ref{gamma-f2}) in powers of $k_\mu$ about the point $x^{\LCp}$, see P2. The leading order of this expansion is
\be
	-  G_{\sigma\sigma} k \!\cdot\! F(x^\LCp) \!\cdot\! \epsilon + \frac{1}{4} \epsilon \!\cdot\!  F(x^\LCp) \!\cdot\! k + 	\ldots = \frac{1 - 3}{12} k \!\cdot\! F(x^\LCp) \!\cdot\! \epsilon + \ldots \;,
	\label{spinscal}
\ee
in which the spin factor gives the same contribution as the scalar part, multiplied by $-3$. From (\ref{spinscal}) we deduce that for scalar QED the tadpole is given by $-1/2$ of (\ref{GSPINKLAR}). Using this to compute the dumbbell for scalar matter immediately confirms that the reducible contribution to the scalar Euler-Heisenberg action is also zero. Again, precisely the same structure was seen in the crossed field case, where we found that $\mathcal{\dot{G}}_{B} - \mathcal{G}_{F} = -2\mathcal{\dot{G}}_{B}$, see (\ref{GbGfXd}). 
%This is in agreement with the constant crossed fields limit that we investigated earlier in this manuscript. 

%Sewing onto \textit{any} diagram with $l$-linear structure $l^\sigma \Delta_{\sigma\rho} \epsilon^\rho$ then gives
%%
%\be\label{gamma-f5}
%	\frac{i e\pi^2}{3} \int\!\frac{\ud^4 l }{(2\pi)^4} \frac{\delta^4(l_\mu)}{l.l}l^\sigma \Delta_{\sigma\rho} \epsilon^{\rho}\epsilon^\mu F_{\mu\nu} \epsilon^\mu l^\nu \int\limits_0^\infty \frac{\ud s}{T}\, e^{-i\tfrac{T}{2}m^2} \bigg|_{\epsilon\epsilon=\eta} = \frac{i e\pi^2}{3} \frac{1}{(2\pi)^4} \frac{1}{4}{\Delta^\nu}_{\rho} {F^{\rho}}_{\nu}  \int\limits_0^\infty \frac{\ud s}{T}\, e^{-i\tfrac{T}{2}m^2}\;.
%\ee

% !TEX root = manuscript.tex
\section{Discussion and outlook}
\label{secConc}

	We have considered one-particle-reducible (1PR) contributions to processes in both constant and non-constant background fields. For the former it was only recently discovered that such contributions could be non-vanishing, in contrast to what had long been assumed in the literature.  We have examined 1PR ``tadpole'' corrections to the two-loop EHL and one-loop propagator in the background in two classes of constant field, Lorentz equivalent to either a constant crossed field ($|{\bf E}| = |{\bf B}|$ and ${\bf E}.{\bf B}=0$) or a pure magnetic field (${\bf E}=0$). In the former case, the tadpole contribution contributes only a divergent factor which can be renormalised away. In the latter case, and in $D = 4$ there is both a divergent renormalisation and finite higher order terms which yield physical corrections to the propagator and, by extension, any other process occurring in a constant magnetic background. These physically relevant corrections have never before been studied to the best our our knowledge.	

	We have also considered background plane waves of arbitrary strength and shape. Here we were able to make a stronger statement; we calculated the 1PR correction to \textit{any} diagram, and showed that this again amounts to a divergence (in $D = 4$) which can be renormalised away. This is consistent with, and goes beyond, one-loop Hamiltonian-picture calculations where the tadpole does not appear due to normal ordering~\cite{Ilderton:2013dba}, as in background-free QED. For all plane waves, including constant crossed fields, we have also confirmed that the dumbbell diagram vanishes identically. Therefore (unlike in the case of magnetic fields) there is no additional two-loop correction to the Euler-Heisenberg effective action coming from the 1PR diagrams. We saw that the reason for this is essentially geometrical -- there is no Lorentz invariant which can be formed which survives the contractions into the field potential demanded by the creation of the dumbbell from the sewing of two tadpoles. That the only part of the tadpole that can see the larger diagram through the mediating photon is linear in the background field is also compatible with the 1PR contribution to the propagator in the crossed field limit being an additional renormalisation.

Our results show that standard lessons from QFT, such as the freedom to ignore tadpoles, does not automatically go over to QFT with background-fields and verifies that the discovery of the 1PR contributions in constant background field QED has physical significance. This holds also for the case in which the photon in Fig.~\ref{figTad} is taken to be an asymptotic state -- as a scattering amplitude this is not zero for a general background, and it describes four wave mixing~\cite{Lundstrom:2005za}, or vacuum emission~\cite{Gies:2017ygp}.

A variety of historic calculations ought now to be revisited with the aim of checking whether the 1PR contributions need to be included to correct the reported result. For example, it has previously been stated that the tadpole diagram vanishes in the combination of a plane wave and \textit{and} constant and homogeneous field~\cite{Fradkin:1991zq}; our results for the magnetic field case demonstrate that this cannot be true. Furthermore, in the combination of a plane wave with any field such that the plane wave is able to contribute to the Schwinger invariants, it becomes clear that there will be a physical contribution \textit{from the plane wave}.  The methods we have developed here will be useful for the future investigation of such cases.
 
	On this note, it would be interesting to extend our results to non-constant magnetic fields and to more realistic models of intense laser fields, and then to examine the physical implications of 1PR corrections. This may allow new insights into particle physics phenomena occurring in terrestrial experiments and astrophysical scenarios such as magnetar environments \cite{Mignani:2016fwz,Capparelli:2017mlv,Turolla:2017tqt,Caiazzo:2018evl}. Another natural extension of this work would be the 1PR contribution to mixed backgrounds such as a constant field accompanied by a plane wave which we anticipate to lead to physical 1PR contributions. Mixed backgrounds have already shown interesting consequences in strong field QED such as boosting the pair production rate, see \cite{schutzhold2008dynamically,Torgrimsson:2017cyb}.  
	Moreover, studying the 1PR contribution to the self-energy in a magnetic background in the strong field limit could be significant in the context of the well-known Ritus (effective) mass shift \cite{RitusShift} and its leading asymptotic behaviour. We would also like to compare 1PR contributions with their known 1PI counterparts. One could also consider the contribution of different types of particles (with different couplings to the background) running in the loop, in order to examine the relative contribution of 1PI and 1PR contributions in BSM models, for example. These topics will be pursued elsewhere.

\acknowledgments
%This work was completed with the support of a Royal Society Newton Mobility Grant, Number NMG\!\! \textbackslash\!\! R1\textbackslash\!\!\! 180368. The authors thank the organisers of EXHILP 2017 and  LPHYS'18, the Insitituto Superior Tecnico in Lisbon and the University of Nottingham which provided useful discussion on background field QED. We are indebted to Christian Schubert, Antonino Di Piazza, Felix Karbstein and  Sebastian Meuren for encouragement and for many helpful discussions. The work of NA was supported by IBS (Institute for Basic Science) under grant No. IBS-R012-D1. JPE is grateful to hospitality of School of Computing, Electronics and Mathematics, University of Plymouth where much of this work was completed. 
The authors thank Antonino Di Piazza, Ralf Sch\"utzhold, Felix Karbstein, Sebastian Meuren and Greger Torgrimsson for useful discussions, and are indebted to Christian Schubert for encouragement, fruitful conversations and helpful comments on this manuscript. JPE is grateful for the hospitality of the Centre for Mathematical Sciences, University of Plymouth where much of this work was completed. JPE and AI are supported by the Royal Society, Newton Mobility Grant NMG\!\! \textbackslash\!\! R1\textbackslash\!\!\! 180368. The work of NA was partially supported by IBS (Institute for Basic Science) under grant No. IBS-R012-D1.

\appendix
% !TEX root = manuscript.tex
\section{Scalar QED}
\label{secApp1}
Here we list the various formulae needed to arrive at the results for scalar QED. Firstly, the scalar EHL has proper time representation \cite{BG0,GKUs1}
\begin{equation}
		\mathcal{L}^{(1)}[F] = \int_{0}^{\infty}\frac{\ud s}{s}(4\pi i s)^{-\frac{D}{2}} \e^{-im^{2}s} \Detp{-\frac{1}{2}}\Big[\frac{ \sinh\Zz}{\Zz} \Big]\,,
	\label{L1WLsc}
\end{equation}
whilst the tree level propagator in a constant background has integral form (in momentum space) \cite{mckeon1994radiative,ScalProp}
\begin{equation}
	D(p | F) = i\int_{0}^{\infty}\!\ud s\, \e^{-im^{2}s}\Detp{-\frac{1}{2}}\Big[\cosh \Zz\Big]\e^{-is p \cdot \frac{\tanh \Zz}{\Zz} \cdot p}\,.
	\label{DpWl}
\end{equation}
With the covariant formulae (\ref{GKEH}) and (\ref{GKProp}) we get the 1PR contribution to the two-loop EHL and one-loop self-energy as given in (\ref{L2sc}) and (\ref{D1sc}). To arrive at explicit formula for the functions of $\Zz$ and $\Zzp$ requires a choice of Lorentz frame and we list the results for the cases considered in the main text below.

\subsection{Constant magnetic field}
To arrive at the results obtained in (\ref{L2B}) and (\ref{D1sc-b}) for the scalar case in a pure magnetic field one needs the following additional formulae:
\bear
\dot{\mathcal{G}}_{B} &=&-\left(\cot z -\frac{1}{z}\right)\Fh\,,\\
p\cdot \frac{\sinh \cZ\cdot\cosh\cZ-\cZ}{\cZ^2\cdot\cosh^2\cZ}\cdot\dot{\mathcal{G'}}_{B}\cdot p&=&-\frac{s}{z}\left[\cot z' - \frac{1}{z'} \right] \left[\frac{\tan z}{z}  - \sec^{2}z \right]p_{\perp}^{2}\,,\\
\frac{\tanh\Zz}{\Zz}&=&g_\parallel+\frac{{\tan}z}{z}g_\perp\,,\\
{\rm tr}\Big(\tanh\Zz\cdot\dot{\mathcal{G}'}_B\Big)&=&2{\rm tan}z({\rm cot} z'-\frac{1}{z'}) \;.
\ear

\section{Weak field limit}
Expanding the $s$-integrand of (\ref{igen}) in the background field provides, to linear order in $z$, the following structures
\begin{align}
	(m - \ps)\left[-is p_{\perp}^{2} \left(\frac{\sec^{2}z}{z} - \frac{\tan z}{z}\right)\right] \e^{-is \frac{\tan z}{z} p_{\perp}^{2}}&= -\frac{2iz}{3}s p_{\perp}^{2}(m - \ps) \e^{-i s p_{\perp}^{2}} + \mathcal{O}(z^{3})\,, \\
	\frac{1}{2}(m - \ps)\left[-is p_{\perp}^{2} \left(\frac{\sec^{2}z}{z} - \frac{\tan z}{z}\right)\right] \tan z\, \e^{-is \frac{\tan z}{z} p_{\perp}^{2}} &= \mathcal{O}(z^{2})\,, \\
	\gamma^{[2}p^{1]}\left[-is p_{\perp}^{2} \left(\frac{\sec^{2}z}{z} - \frac{\tan z}{z}\right)\right] \tan z \,\e^{-is \frac{\tan z}{z} p_{\perp}^{2}} &= \mathcal{O}(z^{2})\,, \\
	\frac{1}{2}\gamma^{[2}p^{1]}\left[-is p_{\perp}^{2} \left(\frac{\sec^{2}z}{z} - \frac{\tan z}{z}\right)\right]\tan^{2}z\, \e^{-is \frac{\tan z}{z} p_{\perp}^{2}} &= \mathcal{O}(z^{3})\,, \\
	\sec^{2}z\, \gamma^{[2}p^{1]} \,\e^{-is \frac{\tan z}{z} p_{\perp}^{2}}&= \gamma^{[2}p^{1]}\e^{-i s p_{\perp}^{2}} + \mathcal{O}(z^{2})\,,\\
	\frac{1}{2} \gamma^{[2}p^{1]}[\gamma^{2}, \gamma^{1}] \sec^{2}z \tan z\, \e^{-is \frac{\tan z}{z} p_{\perp}^{2}}&= \frac{z}{2}\gamma^{[2}p^{1]}[\gamma^{2}, \gamma^{1}]\e^{-is p_{\perp}^{2}} + \mathcal{O}(z^{3})\,,\\
	\frac{1}{2}(m - \ps) [\gamma^{2}, \gamma^{1}] \sec^{2}z\,\e^{-is \frac{\tan z}{z} p_{\perp}^{2}} &= \frac{1}{2}(m - \ps)[\gamma^{2}, \gamma^{1}]\e^{-is p_{\perp}^{2}} + \mathcal{O}(z^{2})\,,\\
	\frac{1}{2}\gamma^{[2}p^{1]}[\gamma^{2}, \gamma^{1}] \tan z \sec^{2}z\, \e^{-is \frac{\tan z}{z} p_{\perp}^{2}} &= \frac{z}{2}\gamma^{[2}p^{1]}[\gamma^{2}, \gamma^{1}]\e^{-is p_{\perp}^{2}} + \mathcal{O}(z^{3})\,.
\end{align}
Now using this in the remaining proper time integral we finally get the result reported in (\ref{S1B}).

\section{Reparameterisation invariant path integrals}
The worldline representation of a generic correlation function %the tadpole in a plane wave background 
involves a functional integral over closed trajectories, $x^\mu(\tau)$, with period $1$. % of the form
%\begin{equation}
%\oint	\mathcal{D}x(\tau) \,  e^{iS[x(\tau)]-i \int J \cdot  x(\tau) \, d\tau } \bigg[  1 + i T^2 \int\limits_0^1\!\ud\tau\, \epsilon \cdot F(x(\tau)) \cdot k  \bigg].
%\end{equation}
%where $S$ is the action and $J$ a current (see (\ref{SJ})).
%
We split the coordinates $x^\mu$ into a centre of mass piece $x^\mu_c$ and a fluctuation $y^\mu$, so $x^\mu(\tau) = x^\mu_c + y^\mu(\tau)$ with centre of mass piece $y^\mu$ obeying %string inspired boundary conditions
\be\label{fluc-def}
	\int\limits_0^1\!\ud\tau\, y^\mu(\tau) = 0 \;.
\ee
The reparameterisation-invariant measure over each coordinate is, in these variables~\cite{Polyakov:1987ez,Mansfield:1990tu},
\be\label{varje}
	\oint\!\mathcal{D} x = \sqrt{2s}\int\!\ud x_c \oint\!\mathcal{D} y \;.
\ee
In four dimensions the free path-integral measure obeys the normalisation

\be\label{pathnorm}
	\oint\!\mathcal{D}^4 x \exp\bigg[-\frac{i}{4s}\int\limits_0^1\!\ud\tau\ \dot{x}^2\bigg] = (4\pi i s)^{-2} \int\!\ud^4 x_c \;.
\ee

\subsection{Worldline propagator properties}
On the unit circle, $\tau \in [0,1]$, the second derivative operator, $\partial_\tau^2$, is invertible on the space of fluctuations $y$ as defined in (\ref{fluc-def}), with inverse $G$ obeying ($G_{\tau\tau'} := G(\tau, \tau^{\prime})$ etc.)
\bea
	G_{\tau\tau'} &=& \frac{1}{2}|\tau-\tau'| - \frac{1}{2}(\tau-\tau')^2 - \frac{1}{12} \;, \\
	\partial_\tau G_{\tau\tau'} &=& \frac{1}{2}\text{sign}(\tau-\tau') - (\tau-\tau') \;, \\
	\partial_\tau^2 G_{\tau\tau'} &=& \delta(\tau-\tau')-1 \;.
\eea
$G$ and $\ddot{G}$ are symmetric, $\dot G$ is antisymmetric, so $\dot{G}_{\tau\tau}=0$. The constant in $G$ is fixed by the condition that it has zero c.o.m.~as in (\ref{fluc-def}). %because otherwise we would not be working in an orthogonormal basis.

\subsection{The Spin factor}
The spin factor that arises in spinor QED can be simplified for a plane wave background  by expanding the exponential function in the defining equation (\ref{Spin}). Subsequently extracting the part at $\mathcal{O}(\epsilon)$ leads to an $N-$fold product of the form ($x_{i} := x(\tau_{i})$)
\begin{equation}
	-\frac{i}{4 N!}\tr_{\gamma} \mathcal{P} \left(\frac{s}{2}\right)^{N} \prod_{i = 1}^{N}\int_{0}^{1}d\tau_{i}\, \sum_{j = 1}^{N} \slashed{n}\slashed{a}^{\prime}(x_{1})\slashed{n}\slashed{a}^{\prime}(x_{2})\cdots  \e^{-i l \cdot x_{j}} \left( \slashed{l} \slashed{\epsilon} - l \cdot \epsilon\right) \cdots \slashed{n}\slashed{a}^{\prime}(x_{N})\,,
\end{equation}
where the sum is over the positions, $x_{j}$, of the term involving the photon. At order $N\geq 3$ one finds there are at least two factors of $\slashed{n}$ which can be brought together by anticommuting past $\slashed{a}'$ and using cyclicity of the trace; because $\slashed{n}\slashed{n}=n^2=0$, these terms vanish. The $N=1$ terms vanishes because the matrix structure is traceless. Hence we are left with the second order contribution which after simplification and selection of the piece linear in the photon polarisation takes the form
\begin{equation}
	-\frac{i s^{2}}{16}\tr_{\gamma}\int_{0}^{1}d\tau \int_{0}^{1}d\tau^{\prime} \,\slashed{n}\slashed{a}^{\prime}(x(\tau)) \left(\slashed{l}\slashed{\epsilon} - l \cdot \epsilon\right)\e^{-i l \cdot x(\tau^{\prime})}.
\end{equation}
Computing the trace, it is a simple step to then incorporate this into $\Gamma_{1}$ and write the result in the form (\ref{gamma-red-1}).

\section{A sewing result}
Sewing the tadpole to a larger diagram requires an integral over the momentum of the intermediate photon, whose form for a plane wave background we study here. We let $l^2=0$ and define
\be\label{1A}
\mathcal{K}_{\mu_1 \mu_2 \cdots \mu_n} := \int\!\ud^4 k\, \frac{1}{k^2} \, \delta^4(k - l) \, k_{\mu_1} k_{\mu_2} \cdots k_{\mu_n} \;.
\ee 
Covariance implies that the integral can only contain products of $g_{\alpha\beta}$ and $l_\mu$. Consider a total of $n \geq 1$ factors of $k$ in the numerator of the integrand of (\ref{1A}). Suppose the integral gave a term containing $r>1$ factors of $g$ and $n-2r$ factors of $l$ (symmetrised). Taking the trace over $r$ pairs of indicies would leave a contribution proportional to $l^{n-2r}$. However, taking the same trace in (\ref{1}) produces $(k \cdot k)^{r-1}$ with no denominator, and the integral vanishes since $l^2=0$. Hence $\mathcal{K}$ must be at most linear in the metric. It follows that
\be\label{2A}
	\mathcal{K}_{\mu_1 \mu_2 \cdots \mu_n} := c_n \big( g_{\mu_1\mu_2} l_{\mu_3} \cdots l_{\mu_n}  + \;  \text{symmetrised}\big) + \lambda_n l_{\mu_1} l_{\mu_2} \cdots l_{\mu_n} \;,
\ee 
where the constant $c_n$ is determined by taking the trace. These arguments do not allow us to determine the coefficient $\lambda_n$, but as explained in the text there is nothing to which this can couple so it can safely be ignored. We have the particular cases
\be
\begin{split}
	\mathcal{K}_{\mu} = \lambda_1 l_\mu \;, \qquad 	\mathcal{K}_{\mu\nu } =  \frac{1}{4}g_{\mu\nu} + \lambda_2 l_\mu l_\nu \;.
\end{split}
\ee 
These results are used to deduce that the plane wave tadpole can only contribute something that can be renormalised away.
%
%(In $\mathcal{K}_{\mu\nu}$ there is a term containing only the metric tensor on the RHS because, in this case only, taking the trace over \textit{all} pairs of indices exactly cancels the $l^2$ in the denominator.)

%\input{6App1-old}
\bibliographystyle{JHEP}
\bibliography{bibPlane}
\end{document}